\documentclass[rmp,aps,amsfonts,amsmath,amssymb,nofootinbib,twocolumn]{revtex4}
\usepackage{graphics}
\usepackage{hyperref}
\usepackage{epsfig}
\usepackage{color}
\usepackage{bm}

\begin{document}

\title{A Measure of Monopole Inertia in the Quantum Spin Ice Yb$_2$Ti$_2$O$_7$ }



\author{LiDong Pan$^1$, N. J. Laurita$^1$, Kate A. Ross$^{1,2}$, Edwin Kermarrec$^3$, Bruce D. Gaulin$^{3,4,5}$, and N. P. Armitage$^1$\\
\medskip
$^1$ Institute for Quantum Matter, Department of Physics and Astronomy, Johns Hopkins University, Baltimore, Maryland 21218, USA\\
$^2$ NIST Center for Neutron Research, NIST, Gaithersburg, Maryland 20899, USA \\
$^3$ Department of Physics and Astronomy, McMaster University, Hamilton, Ontario L8S 4M1, Canada \\
$^4$ Brockhouse Institute for Materials Research, McMaster University, Hamilton, Ontario, L8S 4M1, Canada \\ 
$^5$ Canadian Institute for Advanced Research, 180 Dundas St. W, Toronto, ON, M5G 1Z8, Canada}

\date{\today}

\maketitle

\textbf{An important and continuing theme of modern solid state physics is the realization of exotic excitations in materials (e.g.  quasiparticles) that have no analogy (or have not yet been observed) in the actual physical vacuum of free space.  Although they are not fundamental particles, such quasiparticles do constitute the most basic description of the excited states of the "vacuum" in which they reside \cite{Lacroix11, Balents10, Gingras14}.  In this regard the magnetic textures of the excited states of \textit{spin ices}, magnetic pyrochlore oxides with dominant Ising interactions, are proposed to be modeled as effective magnetic charge monopoles \cite{Ryzhkin05, Castelnovo08}.  Recent inelastic neutron scattering experiments have established the pyrochlore material Yb$_2$Ti$_2$O$_7$ (YbTO) as a quantum spin ice, where in addition to the Ising interactions there are substantial transverse terms that may induce quantum dynamics and - in principle - coherent monopole motion \cite{Ross11a, Applegate12}.   Here we report a combined time domain terahertz spectroscopy (TDTS) and microwave cavity study of YbTO to probe its complex dynamic magnetic susceptibility.  We find that the form of the susceptibility is consistent with monopole motion and a magnetic monopole conductivity can be defined and measured. Using the unique phase sensitive capabilities of these techniques, we observe a sign change in the reactive part of the magnetic response.   In generic models of monopole motion this is only possible through introducing inertial effects, e.g. a mass dependent term, to the equations of motion.  Analogous to conventional electric charge systems, measurement of the conductivity's spectral weight allows us to derive a value for the magnetic monopole mass, which we find to be approximately 1800 electron masses.  Our results establish the magnetic monopoles of quantum spin ice as true coherently propagating quasiparticles of this system.}

\bigskip

Quantum spin ice has received considerable recent attention in the search for quantum spin liquids, as a possible realization for this novel state of matter in which highly quantum entangled spin degrees of freedom evade conventional long range order down to the lowest temperatures \cite{Gingras14}. Magnetic pyrochlore oxides, in which magnetic ions sit at the vertices of corner sharing tetrahedra, provide a three dimensional system where spin ice states are found, when provided with appropriate spin interactions and anisotropy \cite{Gardner10}.  In classical spin ice materials, such as Dy$_2$Ti$_2$O$_7$ and Ho$_2$Ti$_2$O$_7$, large classical spins are forced by strong crystal field anisotropy in the local $<$111$>$ direction, with a primarily ferromagnetic Ising spin interaction. The resulting ground state obeys the Bernal-Fowler ice rules, where each tetrahedron adopts the so-called ``two-in, two-out" configuration.  This is equivalent to the proton configurations in water ice (two-close, two-far) and hence the classical spin ices are found to similarly possess an extensive low temperature residual entropy.  Flipping a single spin (e.g, a dipole excitation) in the spin ice  creates a pair of magnetic monopoles in neighboring tetrahedra, which can then subsequently be separated by additional spin flips resulting in deconfined monopoles \cite{Ryzhkin05, Castelnovo08}.

\begin{figure*}
\includegraphics[trim = 10 10 10 10,width=9.8cm]{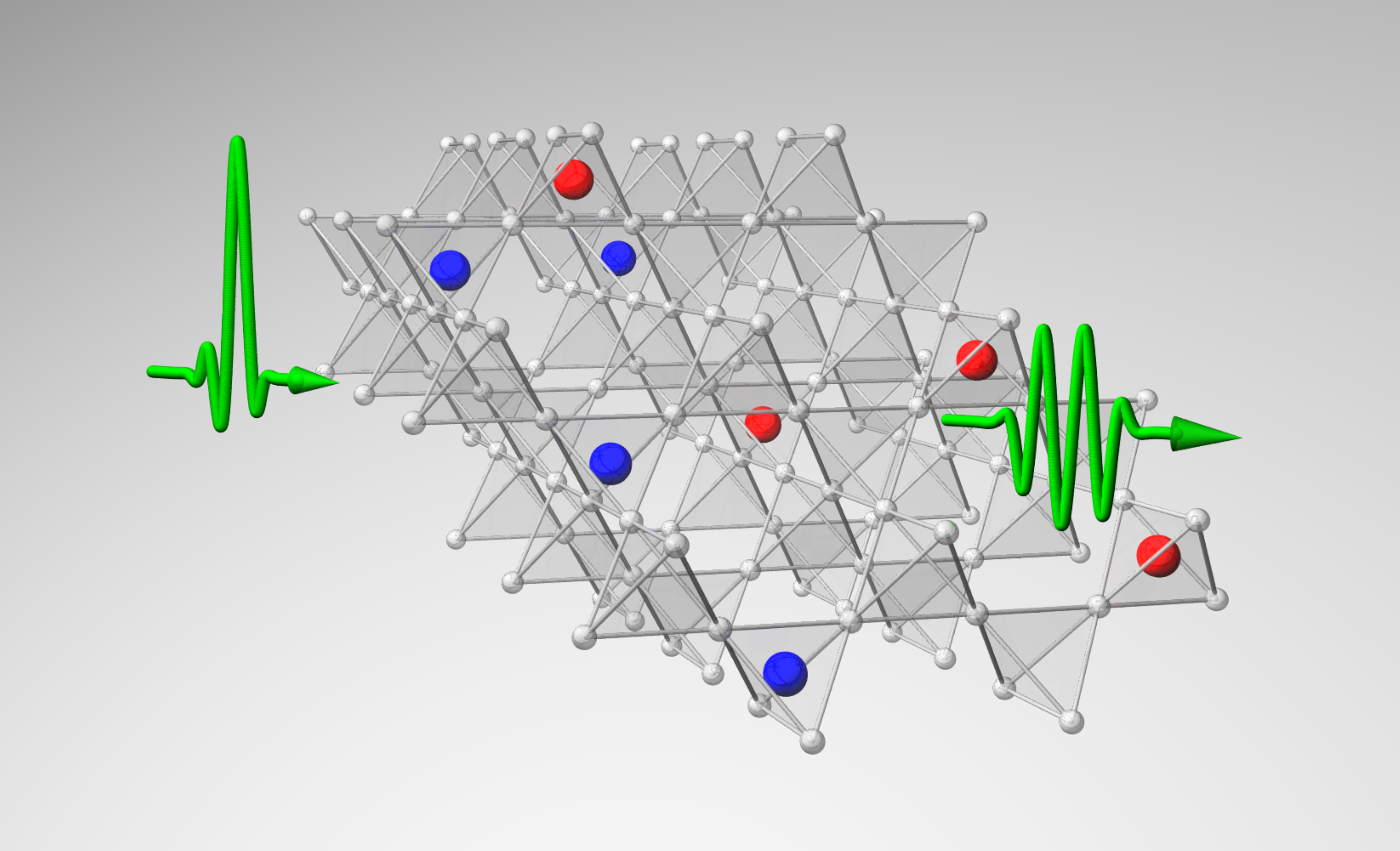}
\label{fig1}
\caption{\textsf{\textbf{Schematic of spin ice and experimental setup}}  Schematic illustration of a quantum spin ice with deconfined monopole and antimonopoles (shown as red and blue spheres), together with incident and transmitted terahertz pulses. In the pyrochlore lattice, magnetic rare earth ions sit at the vertices of corner-sharing tetrahedra, which are shown as grey spheres in the figure.}
\end{figure*}

The slow dynamics of the monopoles in the classical spin ices are still a subject of investigation, but they are believed to be driven by the strong fluctuating transverse component of the dipolar field arising from thermally fluctuating neighboring spins\cite{Ehlers03}.  In contrast, in materials like YbTO, with the addition of finite transverse terms in the spin Hamiltonian, monopole dynamics become inherently quantum and change the situation dramatically \cite{Wan12, Hao14, Savary13}.  The exchange interaction parameters for the magnetic pyrochlore oxide YbTO have been obtained by recent inelastic neutron scattering (INS) at high magnetic fields \cite{Ross11a}, and later confirmed by other experimental and numerical studies \cite{Applegate12, DOrtenzio13, Pan14}.  While the largest interaction is Ising and ferromagnetic, significant non-Ising contributions exist.  The crystal field structure of the material allows the well isolated ground state Kramers doublet to be treated as an effective spin 1/2 moment in the low energy sector \citep{Hodges01, Malkin04}.  These results establish YbTO as an exchange quantum spin ice. Although the exact nature of its ground state is under debate \cite{Ross09, Thompson11, Ross11a, Ross11b, Chang12, Ross12, Hayre13, DOrtenzio13}, at temperatures above a first order transition (T$_c \sim 260$ mK),  YbTO is believed to be in a spin ice-like state, but one whose dynamics are determined by non-Ising terms, which allow the quantum tunneling of magnetic monopoles between tetrahedra. Despite the advances in understanding its behavior, there has been as of yet no definitive evidence for spin ice like correlations in YbTO. With quantum spin liquids lacking an experimentally verified smoking gun, the characterization of the magnetic monopoles in the quantum spin ice state is a relevant and urgent task.  

Here we study the dynamics of the magnetic response in the quantum spin ice regime of YbTO with combined TDTS (schematic illustrated in Fig. 1) and microwave cavity techniques.  Experiments were performed down to a temperature of 1.4 K, which is well below the peak in the low temperature magnetic heat capacity and within the purported quantum spin ice regime \cite{Applegate12} (and well below the estimated scale of the mean-field T$_c \sim$ 3.4 K \cite{Ross11a}). Low frequency electromagnetic spectroscopies can provide highly detailed and accurate information of the dynamic response of the material in the relevant energy range. With the recent advances of phase sensitive optical techniques such as the ones employed here, \textit{complex} material response function can be directly obtained which brings crucial knowledge to our analysis.   The magnetic response of this material is consistent with a model of massive monopole motion.

\begin{figure*}
\includegraphics[trim = 5 5 5 5,width=6.5cm]{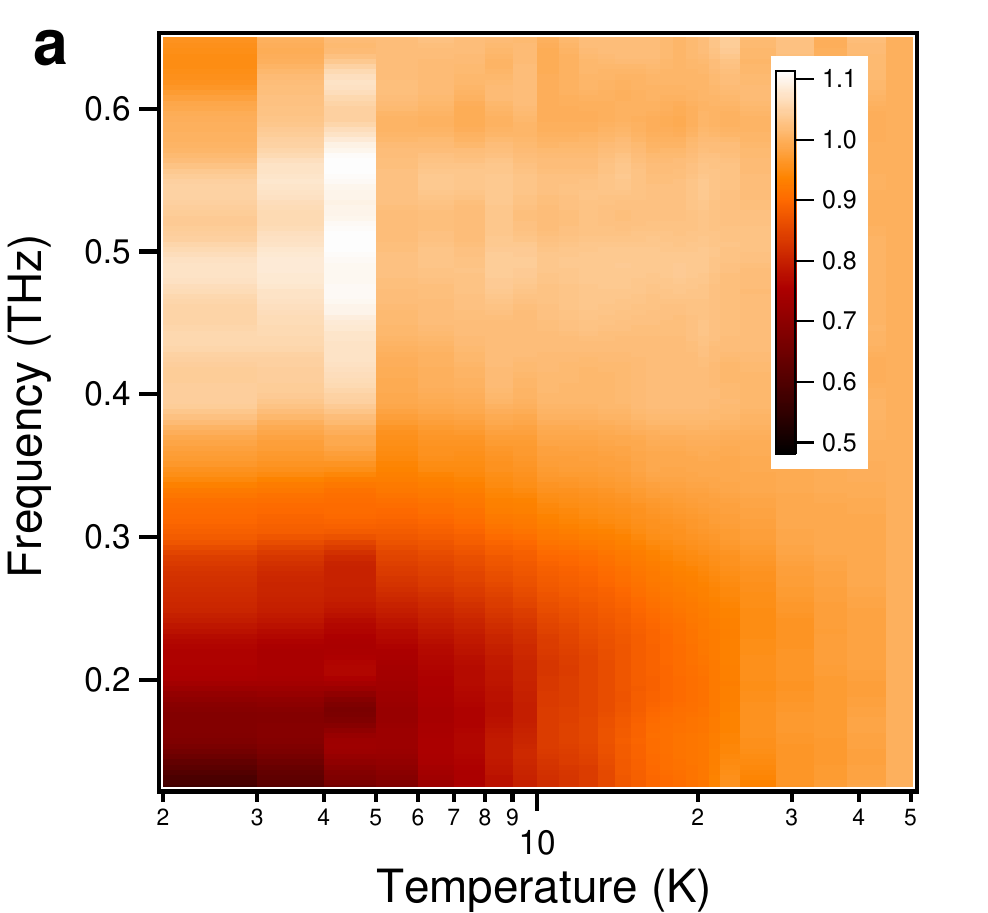}
\includegraphics[trim = 5 5 5 5,width=6.5cm]{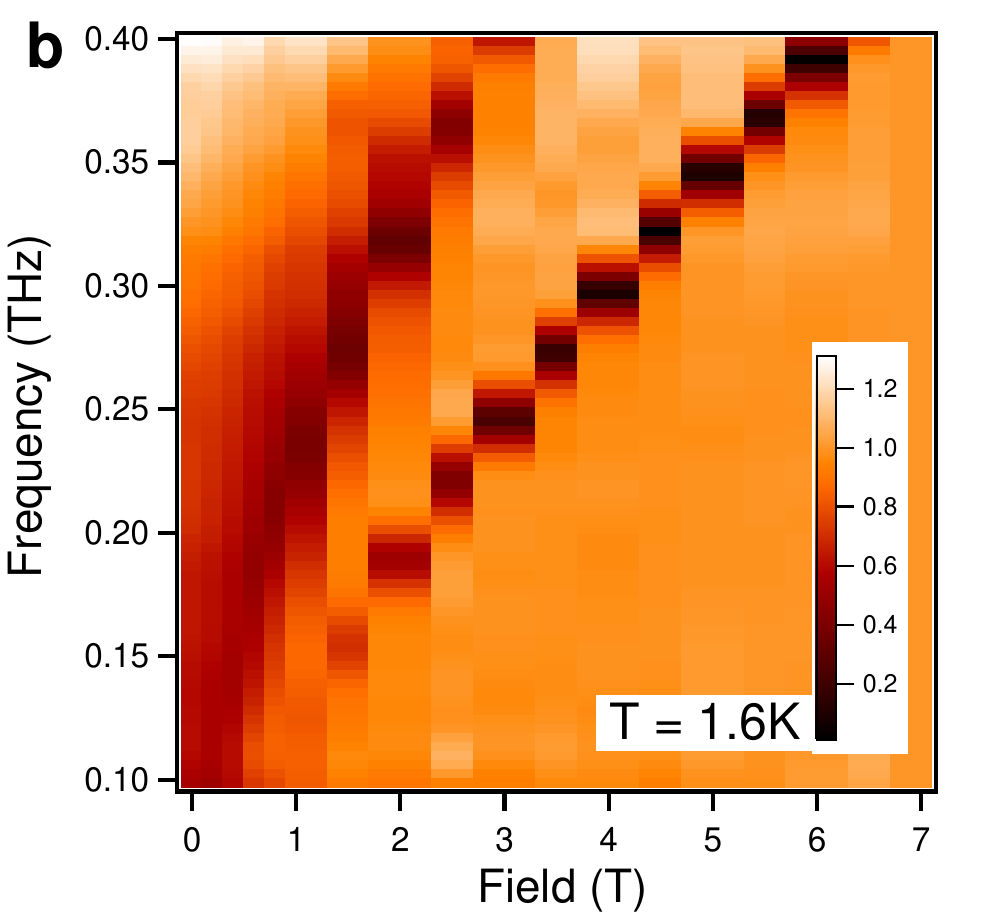}
\label{fig2}
\caption{\textsf{\textbf{Transmission as a function of temperature and field}} (a) The intensity plot of the transmission amplitude from TDTS measured in zero field, as a function of frequency and temperature, normalized with the spectrum at 50 K. (b) Intensity plot of the transmission amplitude as a function of frequency and field, normalized with the spectrum at 7 T. The data in this plot are measured at 1.6 K with \textbf{H}$_{ac}$ $\parallel$ \textbf{H}$_{dc}$ $\perp$ \textbf{k}$_{THz}$ (Voigt geometry). }
\end{figure*}

The physical quantity we directly measure in a TDTS experiment is the complex transmission function $t(\omega)$.  Frequency and temperature dependence of the magnitude of $t(\omega)$ measured from TDTS in zero field are shown in the intensity plot in Fig. 2(a). The spectra are normalized by a scan at 50 K, a temperature well above the onset of magnetic correlations.  An increase in the low energy dissipation is observed as temperature decreases which is unusual for a large gap insulator.  Such an effect is unlikely to come from lattice effects, as the lowest infrared active optical phonon in YbTO has an energy of 2.25 THz.  Moreover, there are no reported dielectric anomalies at low temperature \cite{Malkin04, Vandenborre83}.  The feature is most pronounced for frequencies below 0.4 THz, and for temperatures below 10 K. This trend is reminiscent of the low frequency magnetic susceptibility, as this temperature range coincides with the onset of spin correlations in YbTO and the eventual crossover to the quantum spin ice state. The energy scale of this feature matches the scale of the spin interactions obtained from INS \cite{Ross11a}.

To further demonstrate the magnetic nature of the low frequency absorption, we show the magneto-optical TDTS spectra of YbTO in the Voigt geometry in Fig. 2b. The data shows the THz transmission amplitude as a function of frequency and applied magnetic field taken at 1.6 K, normalized by the spectrum at 7 T.  The field dependence of the absorption features proves the magnetic origin of the zero field dissipative mode.  The behavior of this system in magnetic field have been studied in detail recently \cite{Ross11a,Ross09,Pan14}.  At high fields, YbTO is in a spin polarized state, where excitations are magnon and two-magnon excitations (two dark lines diagonally crossing Fig. 2(b)). As the field decreases, those dipole excitations crossover to weakly confined quantum string-like excitations that connect monopole - antimonopole pairs, which is signified by their increasing effective g-factors below 3 T \cite{Pan14}.  Eventually the magnetic resonance modes lose their sharp structure and the spectra exhibit a broad, diffuse feature at zero field which is consistent with the magnetic monopoles being deconfined in the zero field quantum spin ice state.   We show below that, despite its apparently broad profile, this absorption exhibits features that can be connected to coherent monopole transport.

Given the separation of frequency and temperature scales of the magnetic and lattice degrees of freedom, we can extract the complex magnetic response from our complex transmission.  The spectra at low temperatures are referenced to a scan taken above the onset temperature of the magnetic correlations. More specifically, we have $\mathrm{ln}[t(T)/t(T_{ref})] = in'\omega\chi(\omega)d/2c$, here $\chi(\omega)$ is the complex dynamic magnetic susceptibility of the system in the limit  \textit{q $\rightarrow$ 0}, \textit{c} is the speed of light, and \textit{d} is the sample thickness.  We make the reasonable assumption for an insulator at low temperature and frequency that the index of refraction $n'$ from dielectric effects is real, with little temperature and frequency dependence below 20 K in the THz range.  Further details of the analysis are included in the supplementary information (SI).

\begin{figure*}
\includegraphics[trim = 0 5 5 5,width=8.5cm]{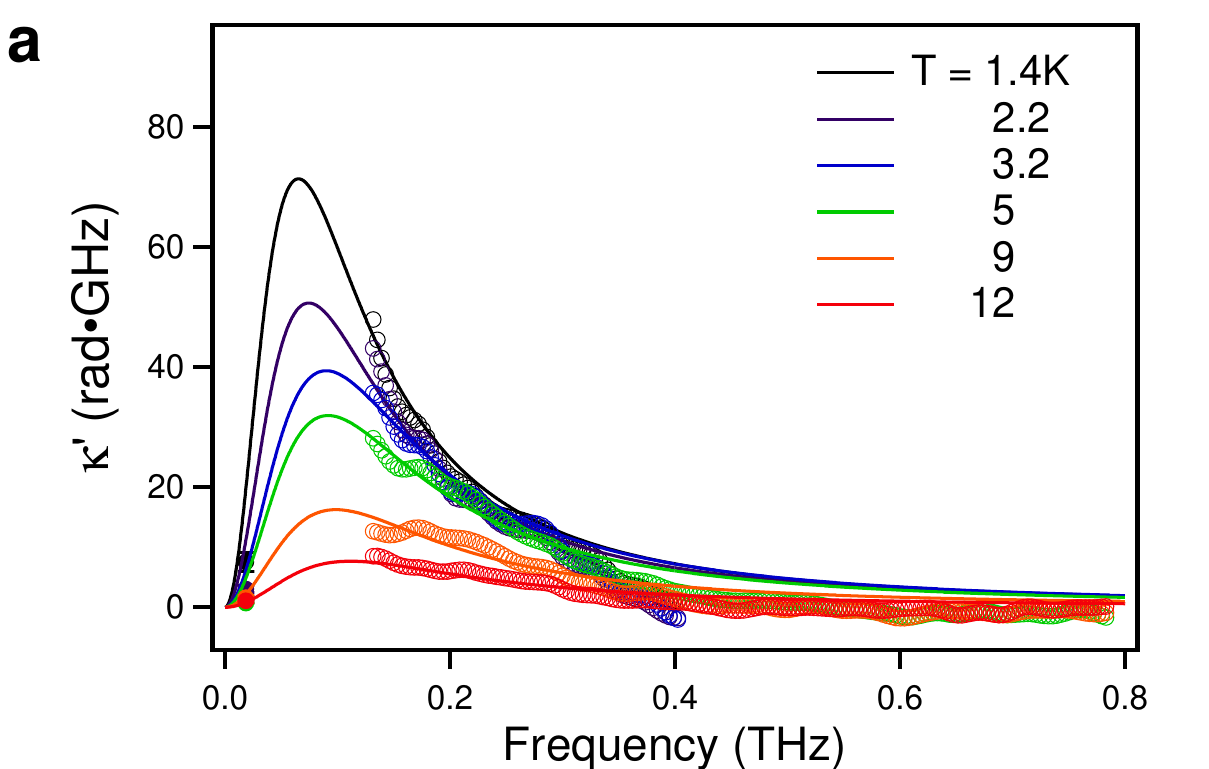}
\includegraphics[trim = 0 5 5 5,width=8.5cm]{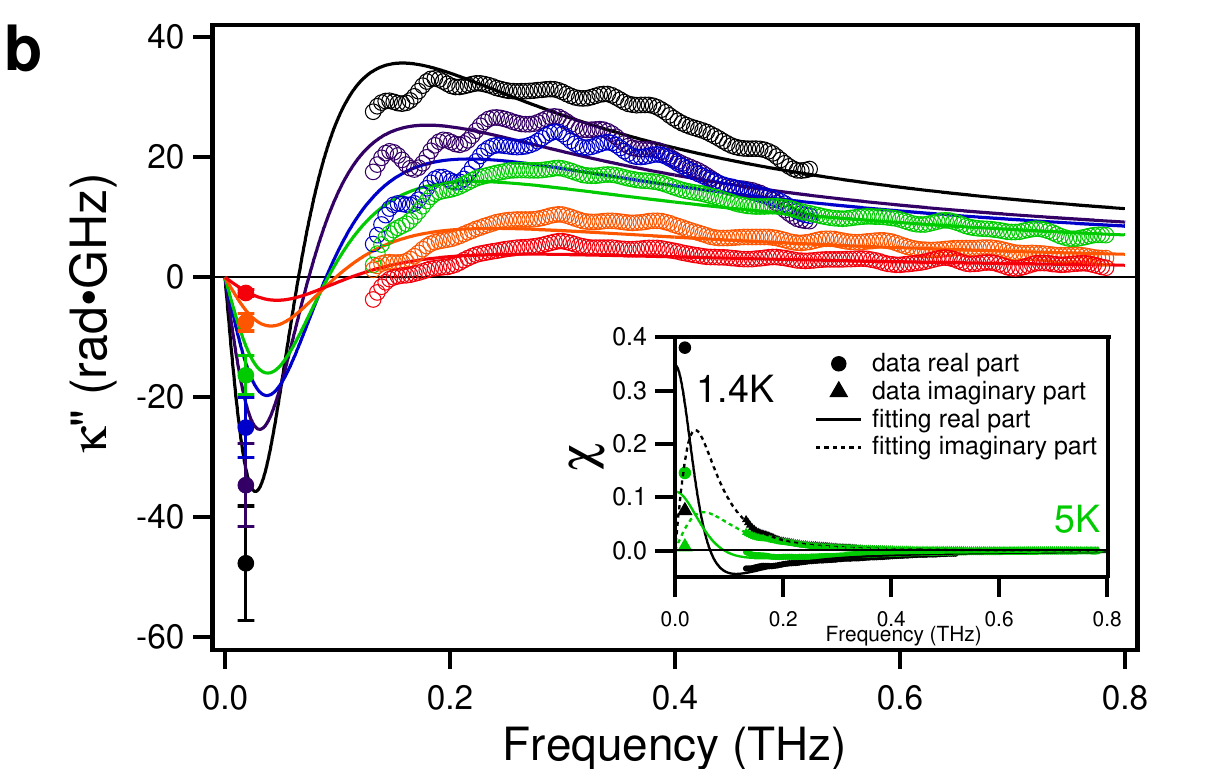}
\label{fig3} 
\caption{ \textsf{\textbf{Real and imaginary parts of measured $\kappa$ with fitting}} Real (a) and imaginary (b) part of the magnetic conductivity $\kappa = \omega\chi(\omega)/i$ as a function of frequency at several temperatures. Solid symbols are data obtained with microwave cavity measurements, while open symbols show data from TDTS. Solid lines are fitting curves as described in the main text. Inset shows the corresponding data and fitting of the complex magnetic susceptibility at two temperatures.}
\end{figure*}

In analogy with conventional electric charge conductivity, we can define a \textit{magnetic monopole conductivity} as $\kappa = \omega\chi(\omega)/i$. In Fig. 3(a) and (b) we show the real and imaginary parts of $\kappa$ as a function of frequency at several temperatures. Data from TDTS are shown as open symbols.  Other than the increasing dissipative response at low temperature, the most striking feature of the data is the trend of the frequency dependence of the reactive response ($\kappa''$), suggesting a sign change of $\chi'$ at a frequency just below the low frequency cut off for TDTS.  This sign change is further confirmed by the addition of microwave cavity measurements at 18.5 GHz, which are shown in Fig. 3 as solid dots.  The combined zero field measurements point to the existence of a zero crossing of $\chi$' at a frequency slightly lower than 0.1 THz.  In the inset to Fig. 3, we show the corresponding complex magnetic susceptibility at low temperatures.  Also, an external dc magnetic field pushes the zero crossing to higher frequencies where reliable TDTS data can be achieved, providing further support for this feature.  Details of the cavity data analysis and these magneto-optical TDTS results are included in the SI.

The existence of the sign change in the reactive magnetic response function is a remarkable feature by itself.   To describe the dynamics of classical spin ices, Ryzhkin proposed a Debye-like model with an equation of motion written as $\mathbf{J}_i = \mu n_{i,m}(q_{i,m} \mathbf{H} - \eta_i \Phi \mathbf{ \Omega})$ \cite{Ryzhkin05}.  Here \textbf{J} is the monopole current, $\mu$ is the monopole mobility, $n_{m}$ is the monopole density, and \textit{q$_m$} is the monopole charge.  $\eta_i $ is an index that takes on values $\pm 1$ for the charge polarity with the subscript $i$ denoting the monopole polarity. The second term in the Ryzhkin model describes a reaction field which originates as a consequence of the configuration entropy due to the motion of monopoles, which prohibits a true direct monopole current even in the thermodynamic limit. Here $\mathbf{\Omega} = \mathbf{M}/ q_m $ is a configuration vector that is proportional to the system's magnetization while $\Phi = \frac{8}{\sqrt{3}} a k_B T$ is a temperature dependent proportionality constant. The Ryzhkin model describes essentially overdamped monopole motion, and generates a frequency dependent susceptibility that is Debye-like in form.  The model has enjoyed great success in the study of classical spin ice materials \cite{Bovo13}, but fails completely in the attempt to account for the sign change observed in $\kappa''$ in the quantum spin ice YbTO.  Moreover, the Ryzhkin model predicts flat frequency dependence of $\kappa'$ whereas the data decreases at high $\omega$.  The only way to capture the sign change in $\kappa''$ and the fall off at high $\omega$ in $\kappa'$ in any classical model of monopole transport is by the inclusion of the inertial effects i.e. a term which is proportional to the second time derivative of the coordinate of interest.  This can be seen from the fact that the leading dependence of the susceptibility in powers of $1/(-i \omega)$ is demonstrative of the dominant term in the equations of motion at a particular frequency.

\begin{figure}
\includegraphics[trim = 0 5 5 5,width=7cm]{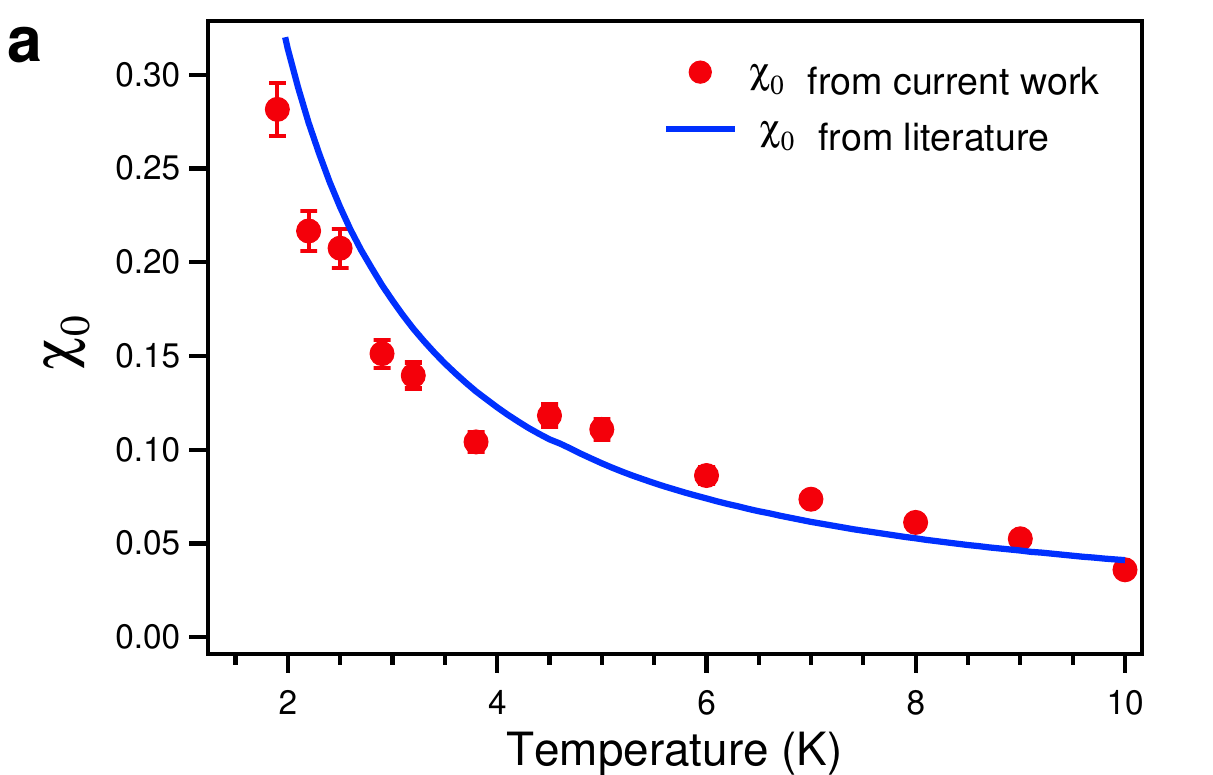}
\includegraphics[trim = 0 5 5 5,width=7cm]{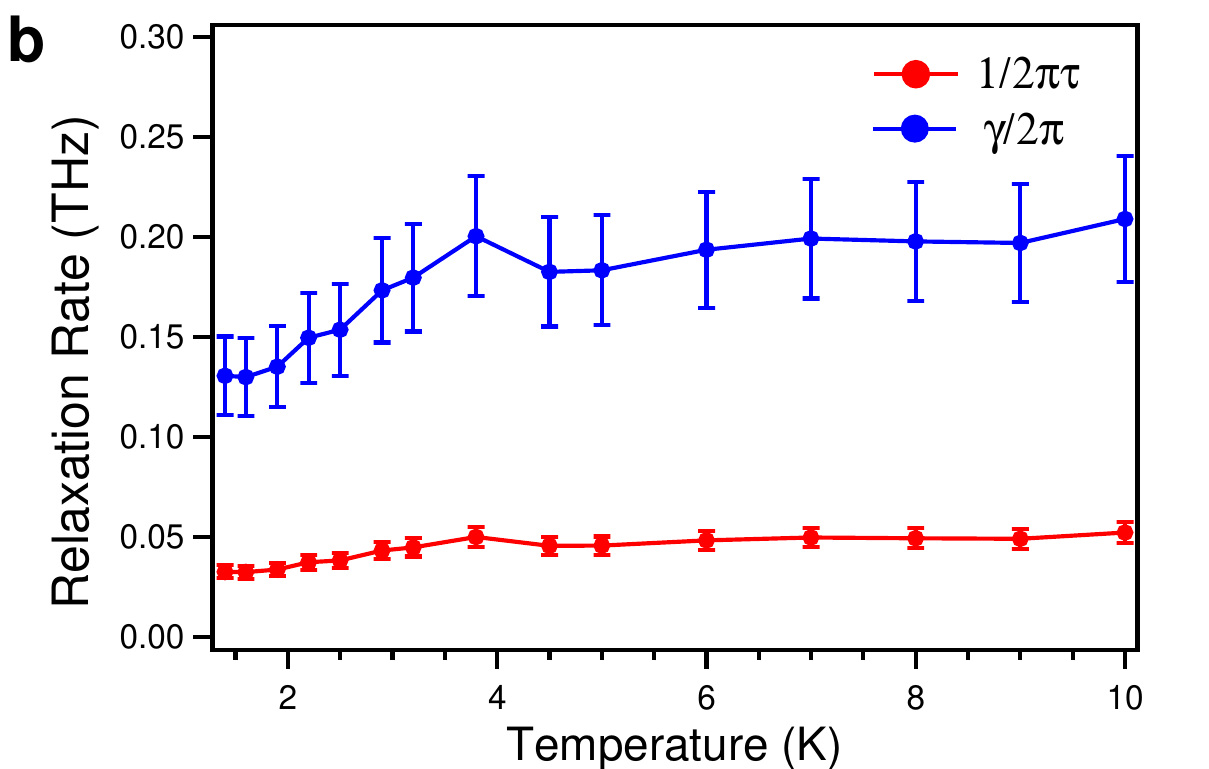}
\includegraphics[trim = 0 5 5 5,width=7cm]{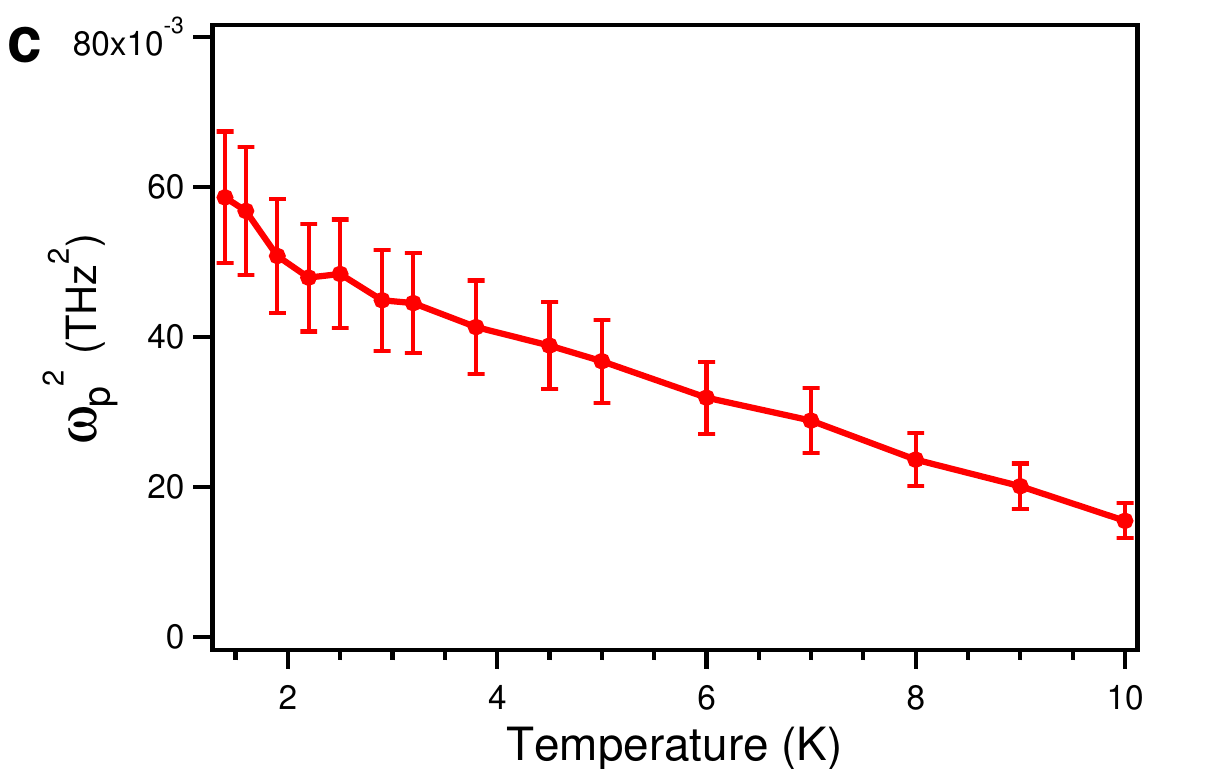}
\label{fig3} 
\caption{ \textsf{\textbf{Temperature dependence of the fitting parameters for the extended Ryzhkin model describe in the main text}}.  (a) $\chi_0$ obtained from the current experiment (dots), and from dc susceptibility measurements reported in  Ref. \cite{DOrtenzio13} (line). (b) Relaxation rates $\gamma$, $\tau^{-1}$.  In this plot, rates are divided by 2$\pi$ to put in units to compare to spectral plots.   To compare these quantities to relaxation in time, quantities should have the 2$\pi$ removed.  (c) Magnetic spectral weight (plasma frequency squared) obtained from fitting.}
\end{figure}

The inclusion of a phenomenological term describing inertial effects in the equation of motion for magnetic monopoles may extend the Ryzhkin model to the quantum spin ice systems.  Our equation of motion then reads $\mathbf{J}_i = \mu n_{i,m}(q_{i,m} \mathbf{H} - \eta_i \Phi \mathbf{ \Omega}) - \dot{\mathbf{J}}_i / \gamma$, where $\gamma$ is a monopole current relaxation rate.  From standard definitions, $ \mu = q_m/ \gamma m_m  $.  Here $m_m$ is an effective monopole inertial mass that arises in the low energy sector through non-Ising exchange that lead to monopole tunneling.  There is no more contradiction inherent in using a classical model to describe a $quantum$ spin ice than there is to use a classical model to describe the conduction of thermally excited charge in a semiconductor where classical inertia also arises through inherently quantum tunneling.  Further motivation for this approach is given in the SI.  Solving in a similar fashion as Ryzhkin yields a Drude-Lorentz style expression for the complex dynamical susceptibility $\chi(\omega)$ 

\begin{equation}
 \chi(\omega) = \frac{q_m^2 / \Phi }{ 1 - i \omega \tau  - \omega^2  \tau/ \gamma}.
 \label{ExtendedRyzhkin}
 \end{equation}

The inertial effects of magnetic monopoles may arise as a consequence of their ability to coherently tunnel between tetrahedra through the finite non-Ising terms in the spin Hamiltonian.   The quantum spin ice system is distinguished by its much lower current damping rate than the classical spin ice systems.  $\kappa(\omega)$ can be fitted to Eq. \ref{ExtendedRyzhkin}, which contains three independent parameters, $\chi_0 = q_m^2 / \Phi$, $\gamma$, and $1/\tau =  n_m \mu \Phi / q_m $.  Here $\chi_0$ is the dc value of the magnetic susceptibility, while $1/ \tau$ is the magnetization relaxation rate that also appears in Ryzkhin's theory. $\gamma$ is a new rate introduced by our treatment that parametrizes monopole momentum relaxation.  Fits shown in Fig. 3 as solid lines are obtained by $simultaneously$ fitting the real and imaginary parts of $\kappa(\omega)$ including both the microwave cavity and the TDTS data. Although the fits are not perfect, it is important to emphasize how restrictive the fitting procedure is (even despite the frequency gap between the TDTS and microwave data) due to the Kramers-Kronig relation between the complex values of $\kappa(\omega)$.  $\kappa'$ at low frequency cannot be changed without affecting $\kappa''$ at higher frequency and vice versa.  This simultaneous fitting of the real and imaginary parts of the data ensures that the essential physics is well captured and constrained.  The ability of this model to capture the salient features provides strong support for the phenomenological description of quantum spin ice developed here, although a more sophisticated model that incorporates a distribution of relaxation times as done recently in the classical spin ices \cite{Bovo13} might potentially improve the fits.   As the spectra themselves evolve continuously with temperature, we have performed these fits up to a reasonably high temperature of 10 K, which is presumably well out of the regime of any spin ice correlations.   As discussed below, in the paramagnetic regime at temperature above the mean-field T$_c$,  we believe that fits to Eq. 1 should be considered only phenomenological without any correspondence of the fit parameters to monopole characteristics.

The temperature dependence of the fitting parameters is shown in Fig. 4. The excellent agreement between $\chi_0$ obtained from the current experiment and the values reported in Ref \onlinecite{DOrtenzio13} with magnetization measurement lends further support to our phenomenological model of the monopole dynamics in quantum spin ice at low temperatures. The satisfying match between the $\chi_0$ obtained from our measurements and the dc value also suggest (due to the above mentioned Kramers-Kronig constraints) there are no significant features at even lower frequency in the magnetic susceptibility of YbTO.  Both relaxation rates $\gamma$ and $1/\tau$ first show a slow decreasing trend upon cooling down to $\sim 4$ K (of order the mean-field T$_c$ \cite{Ross11a}), with a faster dependence below this temperature, before apparent $saturation$ below 2K and down to our lowest measured temperatures.  We believe the data can be interpreted in terms of monopole motion below the mean-field temperature scale.  The magnetization relaxation rate we find at low temperature ($1/ \tau \approx$ 230 GHz) is consistent with the limits put on it by neutron spin echo experiments ($1/ \tau_{echo} \gtrsim $ 250 GHz) \cite{Gardner04}. At this time, the source of the temperature dependence of the momentum relaxation $\gamma$ is unclear.   But in analogy with charge systems, its possible that the temperature independent offset (e.g. a residual resistivity) is caused in part by impurities.  The present system has defects at the 1\% level that are describable in terms of "weak stuffing" \cite{Ross12}, and possible oxygen non-stoichiometry. 

The inclusion of inertial effects also repairs a defect of Ryzhkin's Debye-like model in that it is not, strictly speaking, a fully mathematically consistent magnetic response function as it does not fall off fast enough at high $\omega$ to satisfy the zero temperature first moment sum rule $\int_0^\infty d \omega \;  \omega  \chi_{  \alpha\alpha}''(\mathbf{q}, \omega)  = \frac{1}{2 \pi N \hbar^2} \langle [S_\mathbf{q}^\alpha, [{\cal H}, S_\mathbf{-q}^\alpha]] \rangle$ for local spins \cite{Hohenberg74,Zaliznyak05}. Here  $S_\mathbf{q}^\alpha = \sum_j e^{-i \mathbf{q} \cdot \mathbf{R}_j  } S_{j}^\alpha$ is the Fourier transform of lattice spin operators and ${\cal H}$ is the spin hamiltonian.  A similar violation of a sum rule is a well-known defect of the Debye model in its application to electronic charge relaxation in dielectrics \cite{Onodera93}.   In general the first moment of the $electric$ susceptibility \textit{must} obey the relation $\int_0^\infty d \omega \;   \omega  \chi_e''  =   \int_0^\infty d \omega \;   \sigma'   =     \frac{\pi}{2}  \omega_p^2  $,  where $\omega_p^2 = n_e q_e^2/m_e$ defines the plasma frequency $\omega_p$,  $\sigma$ is the electric charge conductivity, $n_e$ is the electron density and $m_e$ the electron mass.  Debye-like models do not fall off fast enough at high frequency to satisfy this sum rule.  The addition of inertial terms to the equations of motion in Debye-like models gives $\chi$ a high frequency asymptote that goes like $1/(-i \omega)^2$ which insures the first moment's integrability \cite{Onodera93}.  Although at low frequencies in strongly dissipative media inertia can be neglected, at high enough frequencies it must become relevant to satisfy the sum rule.  In electronic charge systems frequent use is made of this sum rule to determine the mass of charge carriers (if the charge density is known).

With the mapping of the form of the monopole conductivity described here to the Drude-Lorentz model, we expect a similar sum rule to exist for the magnetic spectral weight.  To complete the analogy in the present case, we can extract a spectral weight from the fits that is related to a monopole plasma frequency and more fundamental monopole parameters as $\omega_p^2 = n_m q_m^2/m_{m}$.  The temperature dependence of the spectral weight is shown in Fig. 4(c), which illustrates the decreasing trend of the magnetic spectral weight upon warming.  Further analysis of the fits to the susceptibility of the extended Ryzhkin model allows us to extract an effective mass $m_{m}$ of the monopoles from the fitting parameters.  Here we evaluate the low temperature value of $m_{m} = n_m q_m^2\tau/\chi_0\gamma$.  We use the monopole density $n_m$ from the results of Ref. \onlinecite{Applegate12}, which used the Numerical Linked Cluster method to successfully model the heat capacity of YbTO.   Using this computed value for $n_m$, we determine a value of $\sim 1.6 \times 10^{-27}kg$ or $\sim$1800 $m_e$ for the monopole mass at temperatures below 4 K (for further details see SI).

As discussed earlier, the mass of the monopoles originates from their ability to tunnel between sites in the quantum spin ice systems as a consequence of the finite non-Ising exchange terms.  Unfortunately, reliable theoretical estimates for the inertial mass of the monopoles in YbTO in this temperature range do not exist.   However, it is believed that their tunneling rate is governed by the transverse exchange term $J_{z\pm}$ \cite{Applegate12, Hao14}.   We obtain a rough estimate for the monopole mass with a tight-binding model of charges on the diamond lattice (the dual lattice of spins).  In the $ q \rightarrow 0$ limit a calculation gives $m \approx 4 \hbar^2/ (\alpha  J_{z\pm}  d^2)$ where $d$ is the diamond lattice unit cell parameter and $\alpha$ is a temperature dependent dimensionless constant of order unity.   With $\alpha=1$, such a treatment  gives an effective monopole mass of $\sim 2.0 \times 10^{-27}$ kg ($\sim 2200$ $m_e$), which corresponds closely to our experimental value.   As noted above, our spectra evolve continuously into the high temperature paramagnetic regime.   Although one can continue to fit the data to the form of Eq. 1, the fitting parameters become increasingly unphysical above the temperature scale of the mean-field T$_c$.  At 10K one would determine the fitted mass is 10,000 $m_e$ and strongly increasing (see SI), which we do not believe has any physical significance.   This agreement of the fitting parameters with a monopole model at low temperatures and the disagreement at higher temperatures is further testament to the validity of our interpretation.

Following the mass determination of the monopoles and mapping to monopole conductivity, a mobility can also be obtained from an analog of the standard expression $\mu = q_m/m\gamma$, which yields a weakly temperature dependent mobility of $\sim$ 100 ms$^{-1}$T$^{-1}$ for the lowest temperature of our measurement at 1.4 K (a detailed plot can be found in the SI). This value is several orders of magnitude larger than the mobility found in classical spin ice materials \cite{Bovo13}.  This, along with the temperature independent regime of the relaxation rates at low temperature, points to a clear distinction between the monopole transport in the two classes of spin ice materials.  This supports the idea that although in principle the mass would get larger (and inertial effects naively would increase) in the classical Ising limit, quantum tunneling from transverse exchange terms is eventually superseded by incoherent monopole hopping. We note that our TDTS experiments of Dy$_2$Ti$_2$O$_7$ show no temperature or field dependent features in the spectral range reported here. The demonstration of the sign change in the reactive part of the complex magnetic response function is strong evidence of the monopoles' inertial effects in quantum spin ice. With this, their identity as the true coherent quasiparticles of YbTO is further established. 

\section{Methods}

TDTS experiments are performed in a home built spectrometer with two dipole switches as THz emitter and detector. For experiments in magnetic field, a cryogen free superconducting magnet with optical access is used for hosting the sample. Single crystal YbTO samples are grown with optical floating zone method and cut to a disc shape with both surfaces polished parallel and to a mirror finish. The electric field profile of the THz pules after it transmits though the sample and an empty metal aperture are recorded. The ratio of the Fourier transforms of these traces can be further analyzed to give the complex transmission function of the sample, upon which further analyses are carried out.

Microwave experiments are performed by the cavity perturbation technique, whereupon the resonance characteristics (center frequency and quality factor Q) of a superconducting resonator cavity are changed by the introduction of the sample.   These changes are related to the complex generalized susceptibility of the sample under test. In both these techniques the complex susceptibility is measured, which gives both the reactive and the dissipative response of the material.  Further details of these experiments can be found in the SI.


\section*{Acknowledgements}
This work at JHU was supported by the Gordon and Betty Moore Foundation through Grant GBMF2628 to NPA.   The microwave cavity work was supported by the DOE through DE-FG02-08ER46544.  NL had support through the ARCS Foundation. The crystal growth work at McMaster was supported by NSERC. We would like to thank L. Balents, C. Broholm, N. Drichko, M. J. P. Gingras, Z. Hao, S. M. Koohpayeh, J. Lynn, G. Luke, M. Mourigal, M. Robbins, L. Savary, R. R. P. Singh, O. Tchernyshyov and M. Valentine for helpful conversations.

\textit{Competing Interests: }The authors declare that they have no competing financial interests.

\textit{Correspondence: }Correspondence and requests for materials
should be addressed to NPA (email:npa@pha.jhu.edu).

\section*{Author Contribution}

LP performed the THz experiments and data analysis. NL performed the microwave measurements and analysis. KAR, EK, BDG, provided the high quality single crystals.  NPA directed the project.  All authors contributed to discussions on data analysis and writing of the manuscript.

\newpage

\setcounter{figure}{0}
\setcounter{equation}{0}
\setcounter{section}{0}

\newpage

\begin{widetext}

\section*{Supplementary Information}
\subsection{Time-domain THz spectroscopy data analysis for both $\epsilon \neq 1$ and  $\mu \neq 1$}

In a time-domain THz spectroscopy (TDTS) measurement with sample in vacuum, the transmission (neglecting multiple reflections inside the sample) is given by the expression

\begin{equation}
t=\frac{4Z_0Z}{(Z_0+Z)^2}exp(i\frac{(n-1)\omega d}{c}).
\end{equation}

Here $Z = \sqrt{\frac{\mu}{\epsilon}}$ is the wave impedance of the sample, $Z_0$  is the impedance of free space and $n = \sqrt{\mu \epsilon} $ is the complex index of refraction of the sample.  For magnetic insulators, both the dielectric constant and magnetic susceptibility ($\chi_m = \frac{ \mathbf{M}}{ \mathbf{H}}$) may show nontrivial features in the THz range. Therefore the commonly used assumption in the analysis of optical spectroscopy data that $\mu$ = 1 does not apply here. Disregarding this possibility could lead to misassignment of magnetic features as aspects in the charge conductivity \cite{sPilon13}. Here we present a complete treatment of the situation.

Expanding the above equation, we get
\begin{equation}
t=\frac{4\sqrt{\epsilon\mu}}{(\sqrt{\epsilon}+\sqrt{\mu})^2}exp(i\frac{(\sqrt{\epsilon\mu}-1)\omega d}{c}).
\end{equation}
With $\mu = 1 + \chi_m = 1 + \chi'_m + i \chi''_m $ and the reasonable assumption of $\chi_m \ll 1$, this can be rewritten as
\begin{equation}
t=\frac{4\sqrt{\epsilon}(1+\chi_m/2)}{(\sqrt{\epsilon}+1+\chi_m/2)^2}exp(i\frac{(\sqrt{\epsilon}(1+\chi_m/2)-1)\omega d}{c})
\end{equation}

Note that $\epsilon$ and $\chi_m$ are both complex quantities so that both may contribute to optical dissipation and phase retardation. From this equation it is obvious that the principle contribution of the optical response will typically come from the exponential dependence.

As discussed in the main text, a separation of energy and temperature scales for the spin and lattice degrees of freedom allows us to extract the complex magnetic susceptibility from the TDTS measurements for materials like Yb$_2$Ti$_2$O$_7$.  In the present case, the lowest infrared active optical phonon is located around 2.25 THz  and so it is reasonable that the lattice becomes inert for temperatures much lower than this energy scale ($\sim 100$ K) \cite{sVandenborre83}.  Therefore one can assume that the observed low temperature dissipation has a magnetic origin. This assumption is supported by temperature dependence of both the amplitude as well as the phase of the complex transmission function $t$ measured in the TDTS experiments, as shown in Fig. 2(a) of the main text and the Supp. Fig. 1 here.   Here we plot in Fig. 1  the `effective'  index of refraction obtained from TDTS measurement by setting $\mu$ = 1 in the data analysis.  A close inspection of the data reveals that the temperature dependence of the reactive response changes sign at a temperature around 18 K, which is again indicative of the onset of magnetic degrees of freedom at this temperature range.

\begin{figure*}
\includegraphics[trim = 10 10 10 10,width=9.8cm]{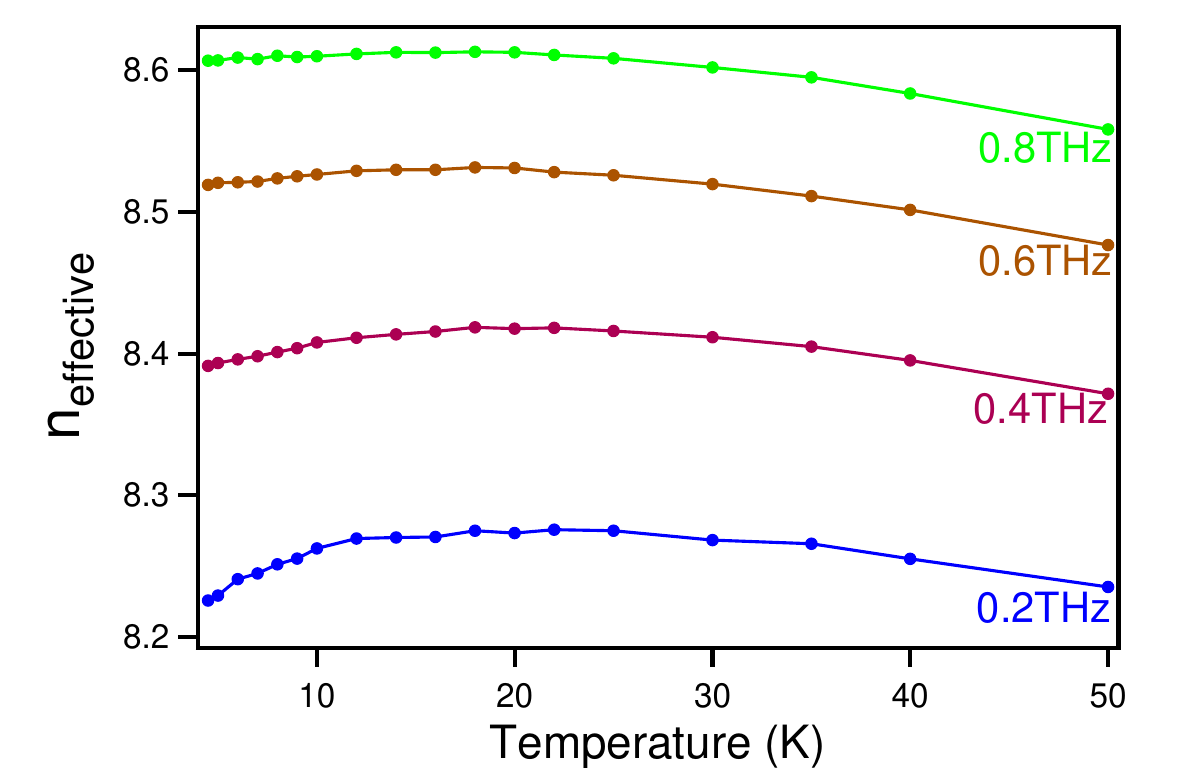}
\label{SI1}
\caption{\textsf{\textbf{Effective index of refraction as a function of temperature}}  Effective real index of refraction obtained from TDTS measurement, when setting $\mu$ = 1 in the data analysis. This data describes the total reactive response from the sample.  Note the deviation from the temperature dependent trend below 15K.   The deviation is more pronounced for lower frequencies.}
\end{figure*}

With this information, then the temperature dependent changes to the optical response at temperatures lower than 18 K can be attributed to a finite value of $\chi_m$ in the THz frequency range, instead of a charge response. To isolate $\chi_m$ from the complex transmission function, we choose a spectrum at $T_{ref}$ = 18 K as a reference spectrum. In the following data analysis, we assume that 18 K is the onset of magnetic response in the spectra, and the dielectric response does not change appreciably for lower temperatures. As discussed above, this assumption is supported by the energies of the optical phonons as well as the features of the TDTS data.

Following the above discussion, we can extract $\chi_m(\omega)$ from the complex transmission via the expression
\begin{equation}
ln(\frac{{t}(T)}{{t}(T_{ref})})=i\frac{n'\chi_m\omega d}{2c}
\end{equation}

Here $n'$ is the real part of the index of refraction of the sample obtained at the reference temperature. With this scheme of data analysis, the magnetic response can be obtained for quantum magnetic materials in suitable systems with similar separation of temperature and energy scales for dielectric and magnetic responses. Here we would like to emphasize that aside from this assumption the treatment presented here is completely model independent, thus it applies to all suitable materials and does not depend on the system being in a particular ground state.

\subsection{Microwave cavity data analysis}

Microwave cavity measurements were performed on several Yb$_2$Ti$_2$O$_7$ samples to measure the complex magnetic susceptibility at a frequency lower than that available in the THz.  As discussed in the main text, these measurements were useful in constraining the Kramers-Kroning consistent fits to the complex susceptibility.  In a typical experiment, one measures the shift in the \textit{complex} resonant frequency $\Delta \widetilde{\omega} = \Delta \omega - i \frac{\Delta \Gamma}{2}$ of a microwave cavity upon the introduction of a sample \cite{sKlein93}.  Here $\Delta \omega = \omega _s - \omega _0$ is the shift in the cavity's resonant frequency and $\Delta \Gamma = \Gamma _s - \Gamma _0$ is the shift in the cavity's bandwidth, or the full width at half maximum of the resonant peak.  The subscripts ``s" and ``0" denote cavity with a sample and empty cavity respectively.  If the introduction of the sample is a small perturbation on the cavity's electromagnetic fields (e.g. the so-called ``adiabatic" regime), it can be shown that the complex frequency shift is directly related to a complex generalized electromagnetic susceptibility $ \zeta$ of the sample,
\begin {equation} 
\frac{ \Delta  \widetilde{\omega}} {\omega _0} = - \gamma_g \zeta
\label{Shift}
\end {equation}
\noindent where $\gamma_g$ is the geometrical factor that depends on the resonance mode of operation and the ratio of sample to cavity volume.  Although the sample's size is usually quite small for such measurements,  a high quality factor of the cavity $Q = \frac{\omega_0}{\Gamma_0} \approx 10^4$ -  $10^8$, allows for an exceptional level of sensitivity.   For details of the method see Ref. \onlinecite{sKlein93}.

In our experiments, we use a superconducting NbTi cylindrical cavity, with T$_c \approx 8.5$ K, $\omega_0 / 2 \pi \approx$ 18.5 GHz, Q $\approx 10^5$, designed to resonate in the TE$_{011}$ mode. The sample is mounted at the magnetic field anti-node at the center of the cavity where the electric field is zero, sitting atop a sapphire rod via a small amount of thermal grease.  The cavity sits at the bottom of a He$_3$ cryostat capable of reaching a base temperature of 500 mK. The generation and detection of the microwave signals are performed by a network analyzer. The microwave response from the cavity is returned and fit to a Lorentzian function to extract the temperature dependent resonant frequency and bandwidth.

In general, the susceptibility ($\zeta$) of Eq. \ref{Shift} is a generalized complex electromagnetic susceptibility which has contributions from both the sample's electric ($\chi_e$) and magnetic ($\chi_m$) susceptibilities the degree of which depends on the sample's position within the cavity.  In this regard it is important to note that, $\zeta$ contains an electric susceptibility contribution even when the sample is placed in a pure magnetic field and vice versa.  As with the TDTS experiments, both contributions must be taken into account.  For small and insulating samples such as Yb$_2$Ti$_2$O$_7$, the electromagnetic fields within the cavity penetrate the sample's volume completely and $k_0 a \ll 1$, where $k_0 = \frac{\omega _0}{c}$ is the microwave wavevector in vacuum and $a$ is a characteristic sample dimension. It can be shown \cite{sZhai00} that in this limit (known as the depolarization regime) and at a magnetic field maximum, the sample's electromagnetic susceptibility can be well approximated by, 

\begin {equation} 
\zeta \approx \frac{\chi_m}{1 + N_m \chi_m} + \frac{1}{10}\frac{\chi_m + 1}{{(1 + N_m\chi_m)}^2}(k_0a)^2 \chi_e
\label{Zeta}
\end {equation}
\noindent where $N_m$ is the demagnetization factor of the sample that depends on sample shape.  Thus in certain limits, electric and magnetic effects may be isolated from each other. 

For the Yb$_2$Ti$_2$O$_7$ sample investigated here, we estimate that $k_0a \approx 0.2$. Similar to the TDTS analysis discussed above, we  assume that changes to the electric and magnetic susceptibilities dominantly occur in two distinct temperature ranges, T $>$ 18 K and T $<$ 18 K respectively.  However, our conclusions are not particularly sensitive to the choice of temperature scale, given that majority of the magnetic susceptibility signal onsets below 10 K.   For T $>$ 18 K, assuming $\chi_m = 0$, plugging Eq. \ref{Zeta} into Eq. \ref{Shift} gives,

\begin{equation}
\frac{\Delta \widetilde{ \omega}} {\omega _0} = - \frac{\gamma_g}{10}  {(k_0a)}^2  \chi _e,  \quad \quad T>18 \text{ K}
\label{ChiE}
\end{equation}
\noindent thus the frequency shift in this temperature range results from the sample's (primarily real) electric susceptibility (polarizability).  Because Yb$_2$Ti$_2$O$_7$ is a good electrical insulator, without any dielectric anomalies at low temperatures,  we can assume that below T = 18 K the real part of $\chi_e$ is constant while the imaginary part is zero.  In the present case, because $\chi_m \ll 1$ and $n \sim \frac{1}{2}$ the second term in Eq. \ref{Zeta} is approximately constant below 18 K, giving

\begin{figure*}
\includegraphics[trim = 5 5 5 5,width=8cm]{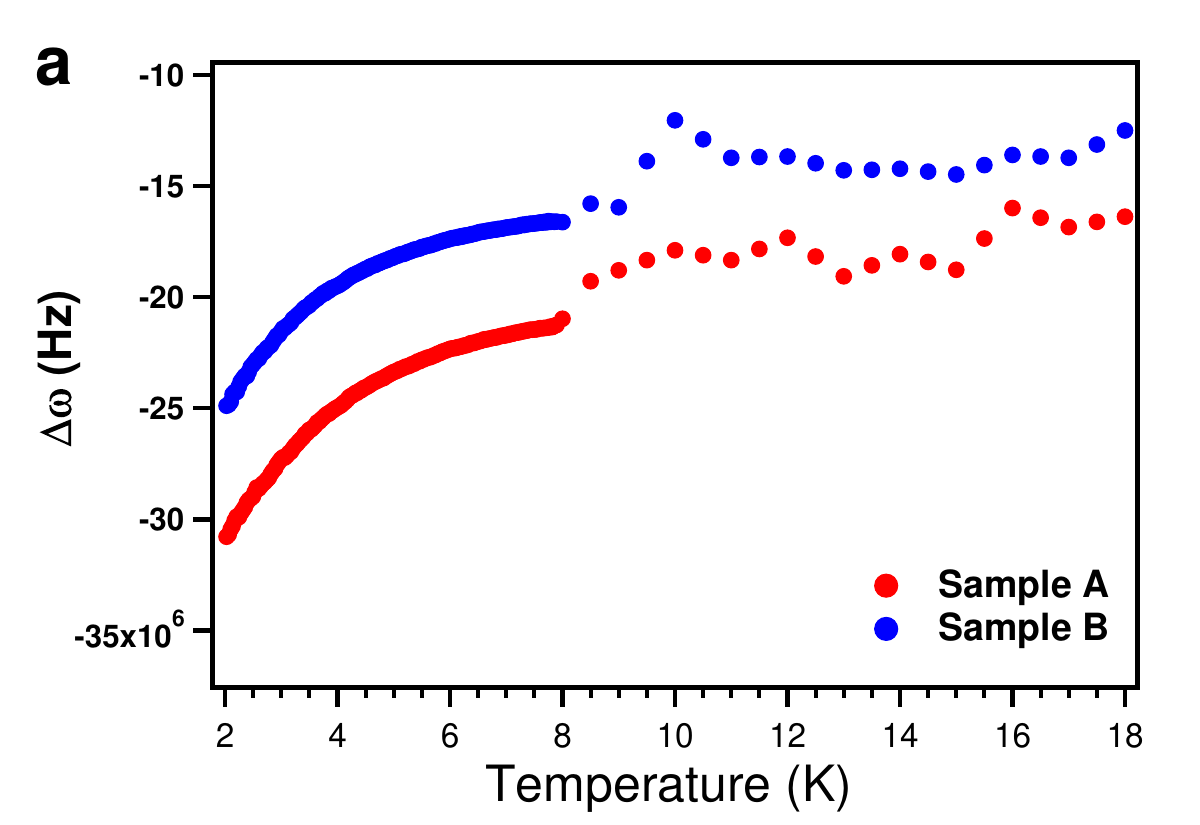}
\includegraphics[trim = 5 5 5 5,width=8cm]{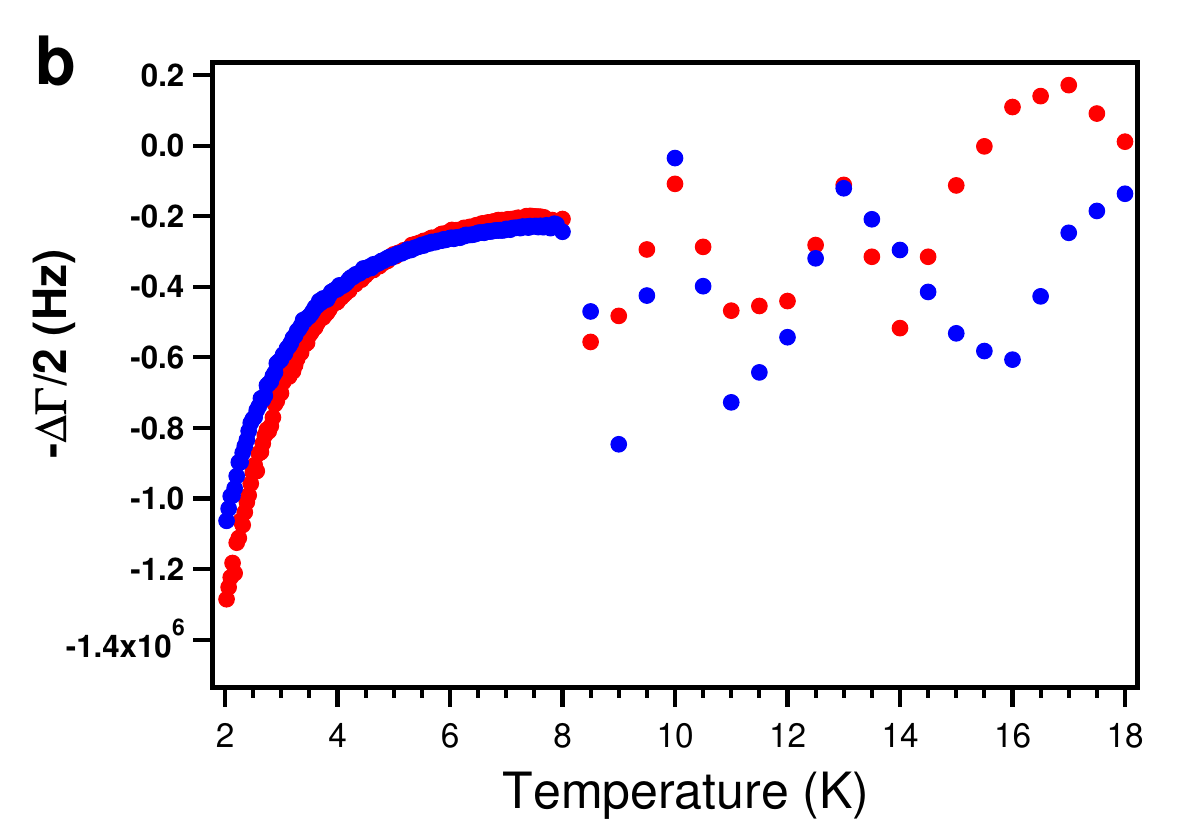}
\includegraphics[trim = 5 5 5 5,width=8cm]{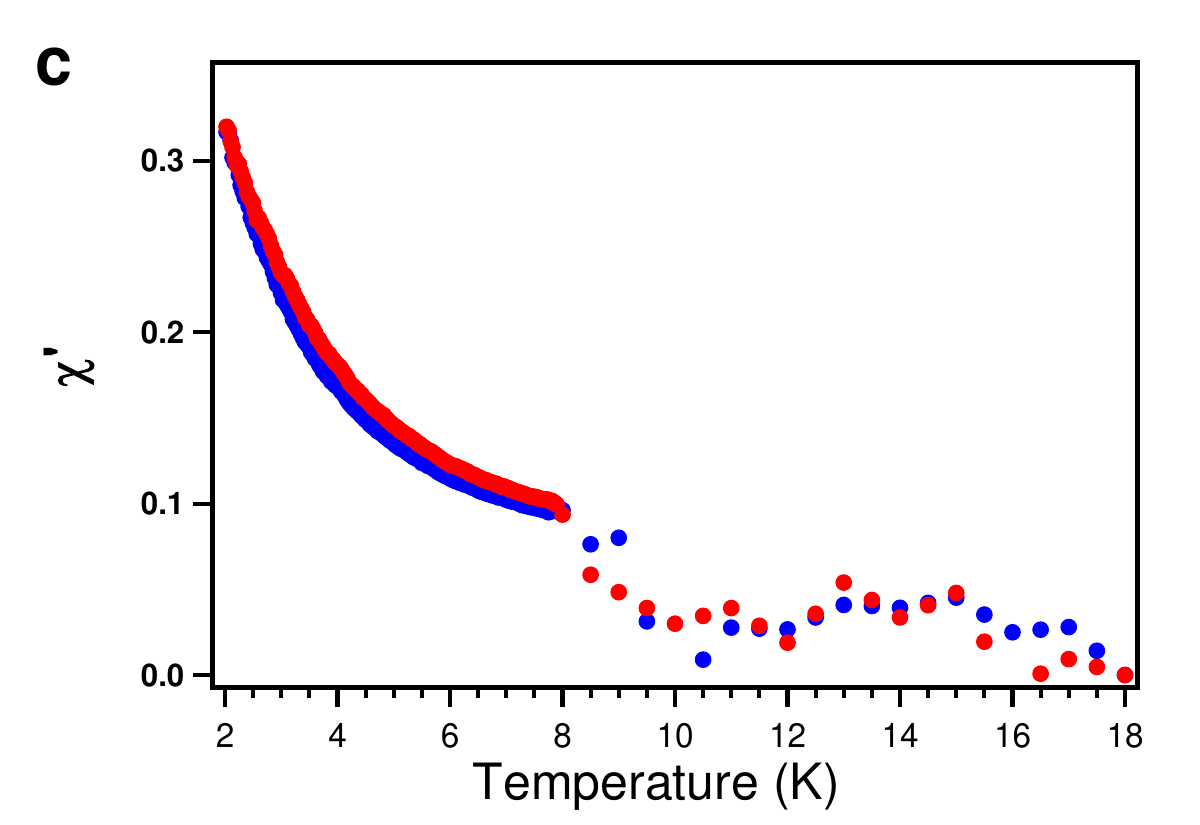}
\includegraphics[trim = 5 5 5 5,width=8cm]{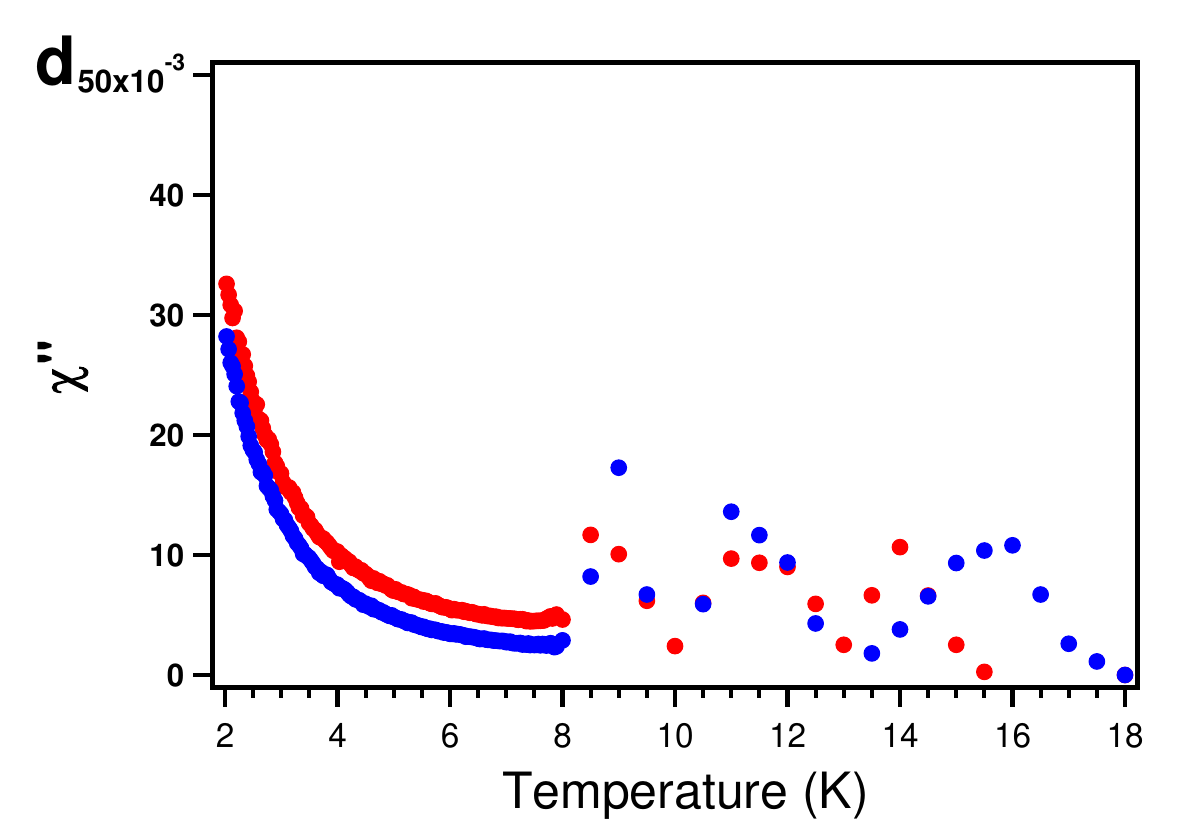}
\label{SI2}
\caption{\textsf{\textbf{Temperature dependence of the microwave cavity data}} Temperature dependence of the real (a) and imaginary part (b) of the complex frequency shift for sample A (red) and sample B (blue), as well as the temperature dependence of the extracted real (c) and imaginary (d) magnetic susceptibility. }
\end{figure*}

\begin{equation}
\frac{ \Delta \widetilde{\omega}} {\omega _0} = - \gamma_g \frac{\chi_m}{1 + N_m \chi_m} + \mathrm{constant}, \quad \quad T<18 \text{ K}.
\label{ChiM}
\end{equation}

\noindent Thus, $\chi_m$ can be extracted using Eq. \ref{ChiM} if $\chi_e$, $\gamma_g$, and $N_m$ are all known.   In our case we extracted the electric susceptibility by linearly extrapolating our THz data into the low frequency regime ($\chi_e$ of 52 and 66 at room temperature and 18 K respectively).   There is about a 17\% rise in $\chi'_e$ from the microwave range to 1 THz.  We then used the shift in resonant frequency from room temperature to 18K of the \textit{loaded} cavity and relate a \textit{change} in $\Delta \omega$ to a \textit{change} in $\chi_e$ over this range to find $\gamma_g$,

\begin{equation}
\Delta (\frac{\Delta \widetilde{\omega}} {\omega _0}) = - \frac{\gamma_g}{10} {(k_0a)}^2 \Delta \chi_e.
\label{ChiE2}
\end{equation}

\noindent The absolute shift between the unloaded and loaded cavity was not used in order to reduce the error caused by a resonant frequency offset that results from removal and replacement of the cavity's bottom plate when inserting a sample.  The demagnetization factors of our samples were estimated from the sample's dimensions.  An extrapolation in frequency of $\chi_e$ from the THz range to the microwave range should be valid in a good electrical insulator like Yb$_2$Ti$_2$O$_7$.  As it was assumed that the magnetic susceptibility of the sample was negligible above 18 K, the magnetic susceptibility below 18 K could then be calculated from Eq. \ref{ChiM} when using the relative shift between the 18 K resonant frequency and the low temperature value.

\begin{table}[t]
\begin{center}
\begin{tabular}{ |c | c | c | c | }
	\hline
\textbf{Sample} & \textbf{Dimensions (mm)} & \textbf{Geo. Factor $\gamma_g$} & \textbf{Demag. Factor $N_m$} \\ 
	\hline
 A & 0.95 $\times$ 1.015 $\times$ 0.582 & 5.6$\times 10^{-4}$ & 0.49\\  
 	\hline
 B & 0.95 $\times$ 1.015 $\times$ 0.479 & 4.9$\times 10^{-4}$ & 0.52  \\
 	\hline
\end{tabular}
\caption{Dimensions, extracted geometrical factors, and calculated demagnetization factors of the two Yb$_2$Ti$_2$O$_7$ samples measured (see text for details.) }
\label{Table:1}
\end{center}
\end{table}

To ensure accuracy of the microwave measurements, two Yb$_2$Ti$_2$O$_7$ samples with different thickness were measured and the results compared to each other. Table \ref{Table:1} summarizes the dimensions, measured geometrical factors, and calculated demagnetization factors of the two samples.  After measuring, sample A was polished to a thickness of 0.479 mm and remeasured as sample B.  Figs. SI2 (a) and (b) show the temperature dependence of the change of the resonant frequency and bandwidth from the two samples respectively.  The superconducting transition of the cavity is seen at T $\approx$ 8.5 K, above which there is significantly reduced resolution.  The reduced resolution above the cavity's transition does not significantly effect our extracted low temperature susceptibility as the signal was small above 8.5 K.  Strong signatures from the sample's magnetic correlations are seen below T $\approx$ 8 K in both the resonant frequency and bandwidth of the loaded cavity. The difference in magnitudes of the shifts from the two samples are caused by different geometrical and demagnetization factors.  However, the general shape of the curves and even more importantly, the direction of the shift is consistent between measurements.  These data and the parameters summarized in table \ref{Table:1} were then used to extract the magnetic susceptibility of each sample.  

In Fig. SI2(c) and (d) we plot the extracted real and imaginary magnetic susceptibility. The data shows a general trend of increase upon cooling, which is what one expects for this material.   We found less than a 10\% difference in the absolute magnetic susceptibilities between the two samples in the temperature range $1.5$ K $<$ T $<$ $8$ K, which should be considered quite good.  The difference could stem from uncertainties in the extracted geometrical factors or in the demagnetization factors resulting from approximating our rectangular prism samples as ellipsoids.

\subsection{Magneto-optical TDTS}

To further investigate the reactive response of the monopole conductivity, we performed a magneto-optical TDTS measurement with \textbf{H}$_{ac}$ $\parallel$ \textbf{H}$_{dc}$ $\perp$ \textbf{k}$_{THz}$ (Voigt geometry). The measurement was conducted with a superconducting magnet with optical access.  The reactive and dissipative response from the measurement at 1.6 K are shown in Fig. SI3 and Fig. 2(b) of the main text, respectively.  Similar data can be found in Ref. \cite{sPan14}.

With the application of an external magnetic field, the diffuse zero field magnetic spectral weight in the dissipative response becomes sharper and is pushed to higher frequencies. Corresponding features are also seen in the reactive response. Here similar data analysis as discussed in the first part of the supplementary information is applied to extract $\chi_m$ in the magneto-optical measurements. As clearly shown in the figure, with the application of magnetic field, the spectra features smoothly shift to higher frequencies with the zero crossing in $\chi'_m$ move to higher frequencies where reliable THz data are available. The data provides strong evidence for the magnetic origin of the main features as well as showing a sign change in the reactive magnetic response. 

\begin{figure*}
\includegraphics[trim = 0 5 5 5,width=12.2cm]{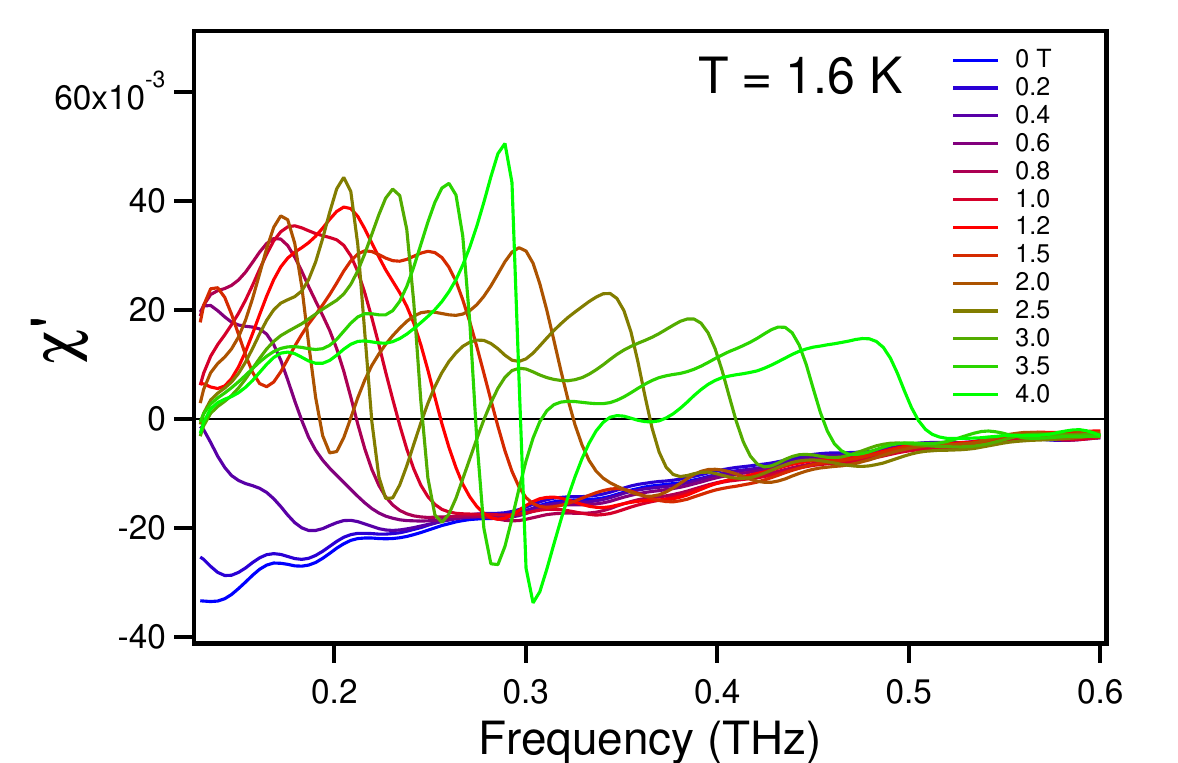}
\label{SI3} 
\caption{ \textsf{\textbf{Real part of $\chi_m$ measured form magneto-optical TDTS}} Real part of $\chi_m$ obtained from a TDTS measurement from Yb$_2$Ti$_2$O$_7$ in field, with temperature at 1.6 K with \textbf{H}$_{ac}$ $\parallel$ \textbf{H}$_{dc}$ $\perp$ \textbf{k}$_{THz}$ (Voigt geometry). With the application of an external field, the zero crossing in $\chi'_m$ is pushed to higher frequencies.}
\end{figure*}

\subsection{Extended Ryzhkin model}

In analogy with previous work done on proton disorder and dielectric relaxation in water ice, Ryzhkin \cite{sRyzhkin05} derived an expression for the magnetic relaxation through monopole motion in classical spin ice.   Starting from the formalism for the local entropy production in irreversible process he derived the monopole flux to be

\begin{equation}
\mathbf{J}_i = \mu n_{i,m}(q_{i,m} \mathbf{H} - \eta_i \Phi \mathbf{ \Omega})
\label{Current}
\end{equation}

\noindent where $i$ refers to positive and negative monopole charge species, $\mu$ is the monopole mobility, $n_{i,m}$ is the monopole density, and $\eta_{i} = \pm 1$.  $q_{i,m}$ is the monopole magnetic charge which is set by the condition that the spin dipole moment $\mu_m $ equals $ q_m a /2 $ where $a$ is spin-spin bond length.   Ryzhkin's expression differs from a simple transport model of free charge by the inclusion of the second term which includes a dependence on a configuration vector  $\mathbf{\Omega}$ that is related to the system magnetization as  $\mathbf{\Omega} = \mathbf{M}/ q_m $.  Finite magnetization reduces entropy and provides a thermodynamic force that opposes the current.   This reaction force originates in the configurational entropy of the monopole vacuum and prohibits a true dc current even in the absence of sample boundaries.  $\Phi$ is a constant of proportionality derived to be $\frac{8}{\sqrt{3}} a k_B T$ in the context of water ice models that retains its relevance here \cite{sRyzhkin97}.   As magnetic relaxation proceeds in a spin-ice through monopole motion the configuration vector is related to the history of the monopole current as 

\begin{equation}
\mathbf{ \Omega}(t) - \mathbf{ \Omega}(0)  = \int_0^t dt' (  \mathbf{ J}_+ - \mathbf{ J}_-).  
\label{Configuration}
\end{equation}

Substituting the Fourier transform of Eq. \ref{Configuration} into Eq. \ref{Current} yields a Debye-like relaxation form for the susceptibility   ($\chi_m = \chi_m' + i\chi_m''$)

\begin{equation}
\chi_m(\omega) = \frac{q_m^2/ \Phi} { 1 - i \omega \tau}
\label{Debye}
\end{equation} 

\noindent with a relaxation time $\tau = \frac{q_m}{ n_m  \mu \Phi } $ in which $n_m$ is the total (thermally excited) monopole density $ n_{+m} + n_{-m}$.  As discussed in the main text, the leading dependence of the susceptibility in powers of $1/(-i \omega)$ is a diagnostic for the dominant term in the equations of motion.   Accordingly Eq. \ref{Debye} is characterized by a $1/(-i \omega)$ fall-off at high frequencies that reflects with a dominant dissipative response.   At low frequencies $\chi_m$ is constant and real showing (e.g. the zeroth power of  $1/(-i \omega)$) that the dominant effect in the dc limit comes from the reaction force.  Note that both real and imaginary parts of Eq. \ref{Debye} will be positive for all frequencies e.g. there is no sign change in $\chi'_m$.

Although such Debye-like relaxation has been found to be a good description of the relaxation processes present in classical spin ice (Ho$_2$Ti$_2$O$_7$ and Dy$_2$Ti$_2$O$_7$ ) at the low (kHz) frequencies used in most experiments,  it is not strictly speaking a fully mathematically consistent response function as it does not fall off fast enough at high $\omega$ to satisfy, for instance, the zero temperature first moment sum rule $\int_0^\infty d \omega \;  \omega  \chi_{ m
 \alpha\alpha}''(\mathbf{q}, \omega)  = \frac{1}{2 \pi N \hbar^2} \langle [S_\mathbf{q}^\alpha, [{\cal H}, S_\mathbf{-q}^\alpha]] \rangle$ for local spins \cite{sHohenberg74,sZaliznyak05}.   Here  $S_\mathbf{q}^\alpha = \sum_j e^{-i \mathbf{q} \cdot \mathbf{R}_j  } S_{j}^\alpha$ is the Fourier transform of lattice spin operators and  ${\cal H}$ is a spin hamiltonian.   The evaluation of the first moment sum rule depends on the details of the exchange interaction, but being equivalent to a double commutator of the spin hamiltonian with spin operators, will yield some finite value.

A similar violation of the sum rule is a well-known defect of the Debye model in its application to electronic charge relaxation in dielectrics \cite{sOnodera93}.   In general, the first moment of the $electric$ susceptibility $must$ obey the sum rule $\int_0^\infty d \omega \;   \omega  \chi_e''  = \frac{\pi}{2} \frac{n_e e^2}{m_e}$ where $n_e$ is the total charge density and $m_e$ the electron mass.   Similar to the Ryzhkin model, the Debye model does not fall off fast enough at high $\omega$ to satisfy the sum rule.  The addition of inertial terms to the Debye model in the classical equations of motion does give a $\chi_m$ a high frequency asymptote that goes like $1/(-i \omega)^2$ which insures its first moment's integrability \cite{sOnodera93}.  Although at low frequencies in strongly dissipative media, inertia can be neglected, at high enough frequencies it must become relevant to satisfy the sum rule.   Frequently in optical studies of charge systems a restricted low energy sum rule is applicable, where the upper limit of integration is taken to be finite, but then $n_e$ is replaced by the density of electrons in the low energy sector, and $m_e$ is replaced by a mass that is renormalized by band and/or interaction effects.

Therefore, for spin ice, irrespective of their exact form, effects beyond Ryzhkin's treatment must become relevant at high enough frequency to satisfy the first moment sum rule.  We have argued in the main text that inertial effects must be included to understand magnetic relaxation in the $quantum$ spin ices.  In a similar fashion to charge in dielectrics, the Ryzhkin expression Eq. \ref{Debye} can be amended by including a phenomenological inertial term to Eq. \ref{Current}.   As discussed in the main text, it will have the form of the final term in the expression

\begin{equation}
\mathbf{J}_i = \mu n_{i,m}(q_{i,m} \mathbf{H} - \eta_i \Phi \mathbf{ \Omega}) - \mathbf{\dot{J}}_i / \gamma 
\label{CurrentInertia}
\end{equation}

\noindent where $\gamma$ is a current relaxation rate.   

From the standard definitions, $q_m/ \mu = \gamma m$.  In what follows, $m$ will be an effective inertial mass which arises in the low energy sector through non-Ising exchanges that lead to monopole tunneling.  There is no more contradiction inherent in using a classical model to describe a $quantum$ spin ice than there is to use a classical model to describe the conduction of thermally excited charge in a semiconductor where classical inertia also arises through inherently quantum tunneling.   Solving in the same fashion as in the above Ryzhkin case gives a classical equation of motion, the terms of which are instantly familiar and map to the form of a damped harmonic oscillator.

\begin{equation}
\mathbf{\ddot{M}} + \gamma  \mathbf{\dot{M}} + \frac{n_m \Phi}{m} \mathbf{M}  = \frac{n_m q_m^2}{m}  \mathbf{H}.
\label{InertiaEquationMotion}
\end{equation}

Solving for the susceptibility we have 

\begin{equation}
\chi_m(\omega) = \frac{q_m^2 / \Phi }{ 1  - i \omega \gamma  m / n_m \Phi  - \omega^2  m / n_m \Phi  }.
\label{SusceptibilityInertia}
\end{equation}

\noindent which in terms of $\tau$ (and $\gamma$) reads

\begin{equation}
\chi_m(\omega) = \frac{q_m^2 / \Phi }{ 1 - i \omega \tau   - \omega^2  \tau/ \gamma }.
\label{SusceptibilityInertia2}
\end{equation}

With appropriate substitutions ($\omega_0^2 =  \frac{n_m \Phi}{m}  $ and $\omega_p^2 =  \frac{n_m q_m^2}{m}$)  Eq. \ref{SusceptibilityInertia} can be seen to be equivalent to the  Drude-Lorentz equations that describe the response of a classical $electric$ charge oscillator. 

\begin{equation}
\chi_m(\omega) = \frac{\omega_p^2  }{ \omega_0^2     - \omega^2  - i \omega \gamma  }.
\label{SusceptibilityInertia3}
\end{equation}

\noindent In the limit where $\gamma \rightarrow \infty$ Eq. \ref{SusceptibilityInertia2} recovers Ryzhkin's expression for Debye-like relaxation.   In this limit the zero crossing in $\chi''$ will move to infinity.

\begin{figure*}
\includegraphics[trim = 0 5 5 5,width=7cm]{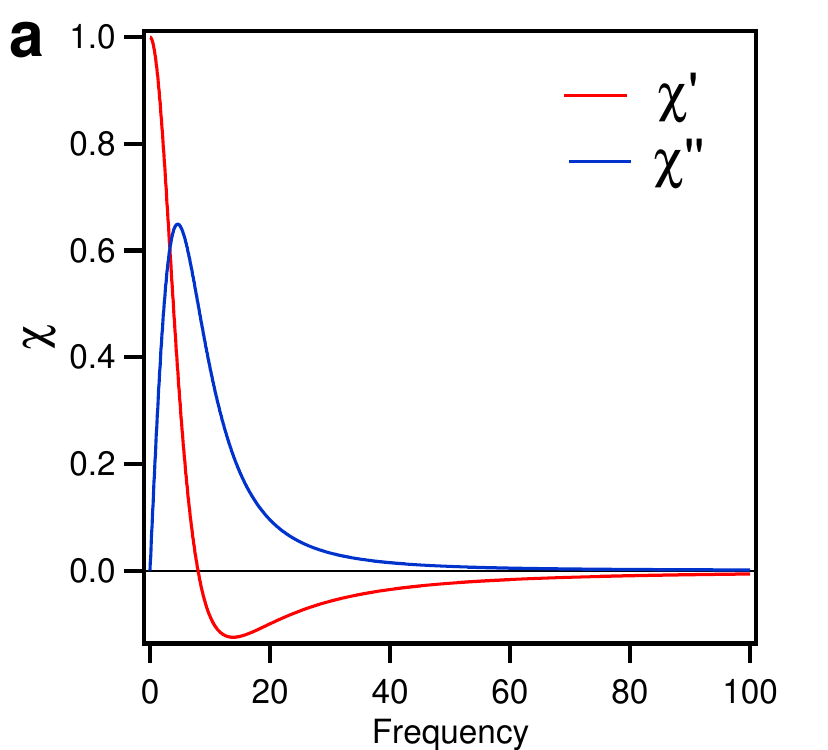}
\includegraphics[trim = 0 5 5 5,width=7cm]{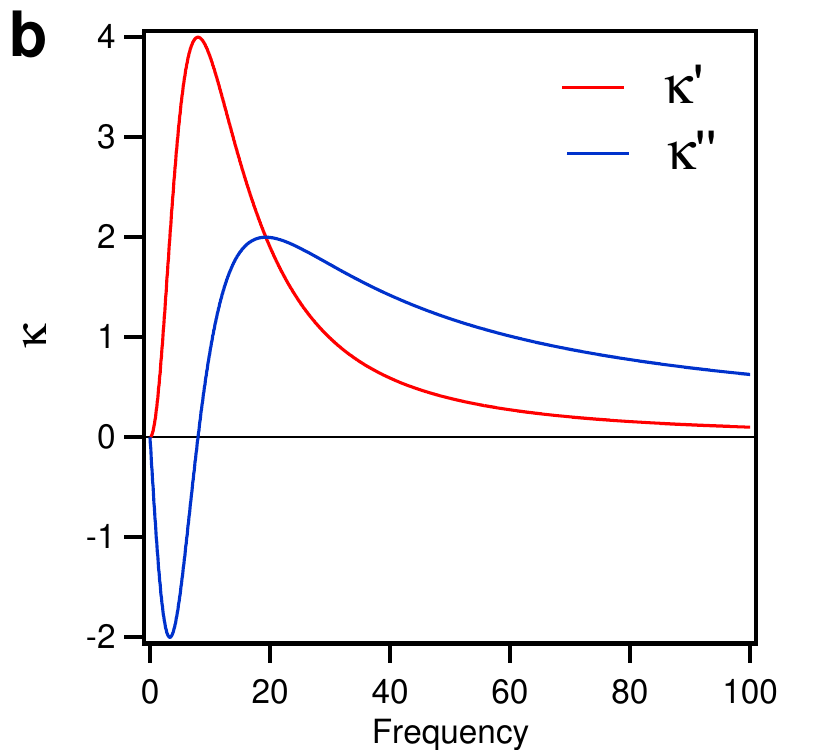}
\includegraphics[trim = 0 5 5 5,width=7cm]{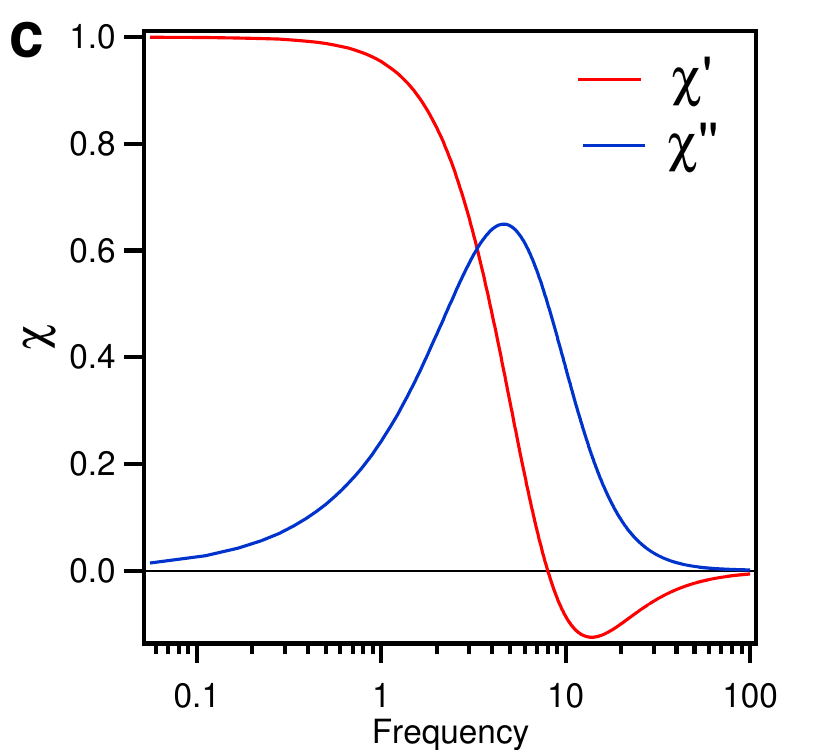}
\includegraphics[trim = 0 5 5 5,width=7cm]{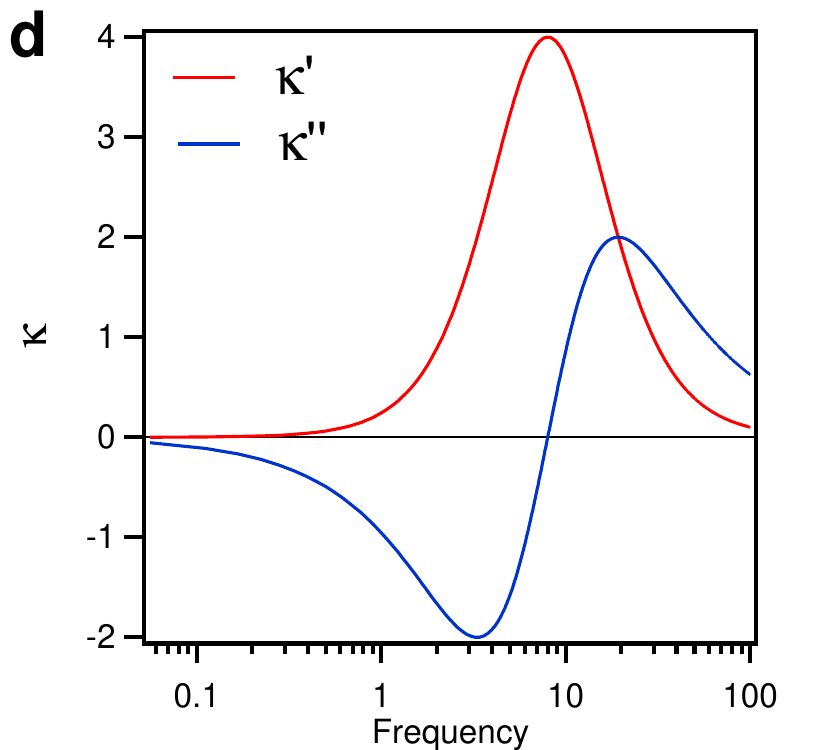}
\includegraphics[trim = 0 5 5 5,width=7cm]{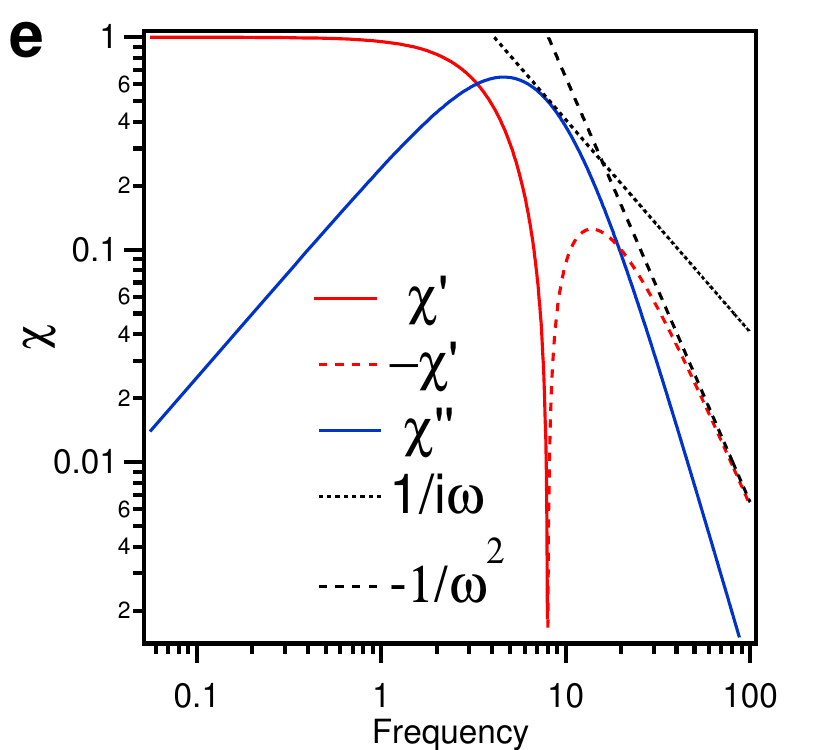}
\includegraphics[trim = 0 5 5 5,width=7cm]{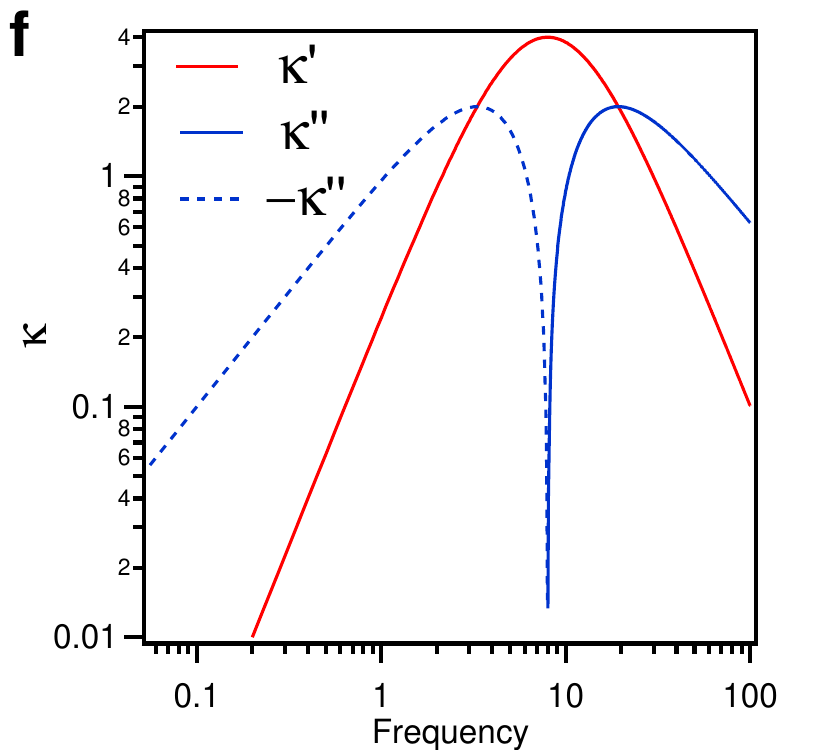}
\label{SI4} 
\caption{ \textsf{\textbf{Simulations of frequency dependent monopole susceptibility and conductivity}}  Plots of the susceptibility $\chi_m$ and the magnetic conductivity $\kappa$ with example parameters $1/ \tau = 4$ and $\gamma = 16$ on both linear, log, and log-log scales.}
\end{figure*}

From Eq. \ref{SusceptibilityInertia2} for the susceptibility and in analogy with charge conductivity we can define a magnetic monopole conductivity $\kappa(\omega) = - i \omega \chi_m(\omega) = \kappa' + i \kappa ''$.   As discussed in the main text, the quantity $\omega \chi_m$ is the quantity directly measured in TDTS experiments.  Due to mapping of our expressions Eqs. \ref{SusceptibilityInertia} and \ref{SusceptibilityInertia2} to the Drude-Lorentz model, the magnetic conductivity must obey a low energy sum rule where 

\begin{equation}
\int_0^{\omega^*} d \omega \;  \kappa'(\omega) =   \frac{\pi}{2} \frac{n_m q_m^2}{m} .
\label{KappaSumRule}
\end{equation}

Here $\omega^*$ is a energy cut-off that must be smaller than a scale on the order of the Ising exchange parameter  $J_{zz}$, but much larger than $\gamma$.   Note that this is extension to the usual first moment sum rule for spin systems as $\kappa = - i \omega \chi_m$.   It is also important to note that this monopole conductivity $\kappa$  is not the same as the spin conductivity which is defined as the response of a spin current to the magnetic field \textit{gradient} in the small momentum limit \cite{sScalapino93a,sAlvarez02a}.

In Fig. SI4 we plot the results of function Eq. \ref{SusceptibilityInertia2} for both susceptibility $\chi_m$ and the magnetic conductivity $\kappa$ for the range of $\tau$ and $\gamma$ we believe is relevant to Yb$_2$Ti$_2$O$_7$.   Here we use example parameters $1/ \tau = 4$ and $\gamma = 16$ on both linear, log, and log-log scales.  These plots highlight the utility of analyzing both of these quantities.  We see that the dissipative part of the susceptibility $\chi''$ peaks near $1/\tau$, which is the usual Debye-like relaxation behavior.  In  $\chi_m$ the scale of $\gamma$ appears only as a subtle change in the high frequency power law, which manifests itself as a change of slope on the log-log plot.  An experimentalist doing typical low frequency susceptibility or neutron scattering experiments (that measure only $\chi''_m$)  would be likely to be completely unaware of the high frequency scale where inertial effects become relevant.   The existence of inertial effects are manifest due to the sign change in $\chi'_m$  (which are absent in the Debye-like Ryzhkin expression).  The magnetic conductivity $\kappa$ exhibits both frequency scales prominently.   The imaginary part of $\kappa$ exhibits a negative extremum at $1/\tau$.   $\gamma$ is seen as the frequency where $\kappa' > \kappa''$ and   $\kappa''$ exhibits a maximum.   The upper right panel of $\kappa$ on linear scale demonstrates that if one is measuring at frequencies well above $1/\tau$ the response will be indistinguishable from a Drude-like transport and appear as a monopole metal.   In that case the width of the peak in the real part of the monopole conductivity would give $\gamma$.   

The lower left panel also demonstrates the point made in the text that the bounding behavior of $\chi_m$ is a power law in $1/(i \omega)$, where the power indicates what term in the classical equation of motion is dominant at a particular frequency.   This is because there are three ``response" terms in the equation of motion, which leads to three terms in the denominator of the susceptibility that can be written as increasing powers in $i \omega$.  In different frequency regimes, different terms on the bottom of Eq. \ref{SusceptibilityInertia2} dominate.  Distinct frequency regions of $1/(i \omega)^0$, $1/(i \omega)$, and $1/(i \omega)^2$ are therefore apparent in the plots of  Fig. SI4.

A few more comments are in order about the applicability of our model for $\chi_m$.  Our approach above was to first define macroscopic variables ($\mathbf{J}, \mathbf{M}$ etc.) and then combine them in a way that was consistent with both an entropic restoring force and inertial effects.  The perceptive reader may make the reasonable objection to this approach that the long time averaging which is necessary to define an entropic force is inconsistent with the short time scales on which inertial effects will become apparent.  An alternative and more rigorous approach would be to \textit{first} write down a Langevin type equation that includes inertia and \textit{then} do the averaging in the Langevin style.  In contrast our approach was one where the order of these operations was essentially reversed.  It is not $a$ $priori$ clear if the steps of including the dynamics as such and then averaging or vice versa commute.  In general they do not.

Of course, this issue is not unique to the present case.  Such considerations are ubiquitous when a high frequency response is considered in the presence of inertial forces (e.g. polymer relaxation, high frequency phonon propagation, etc.).  For instance, related issues arise in the inclusion of inertial effects in the context of dielectric relaxation.   A rigorous Langevin treatment of an assembly of non-interacting fixed axis rotators with finite moments of inertia yields a susceptibility that has the complicated form of a infinite set of continued fractions \cite{sCoffey04} (Similar expressions were obtained earlier by different methods by Gross \cite{sGross55} and by Sack \cite{sSack57}).  However, for small inertia, this expression can be approximated by the first convergent, and is known as the Rocard equation, after Rocard who in 1933 first derived it \cite{sRoccard33a} when including inertial effects in the context of the Debye theory.   The Rocard equation is completely equivalent to the expression we have used above.   This example shows that there are cases and limits in which the ordering in the treatment of including inertia and ensemble averaging does not effect the end result.  Unfortunately, it is not known at this time, to what extent these effects matter or how to perform a Langevin style analysis for dynamics in the spin ices.   This is a very open area for theoretical inquiry.

\subsection{Monopole mass and mobility}

\begin{figure*}
\includegraphics[trim = 0 5 5 5,width=10cm]{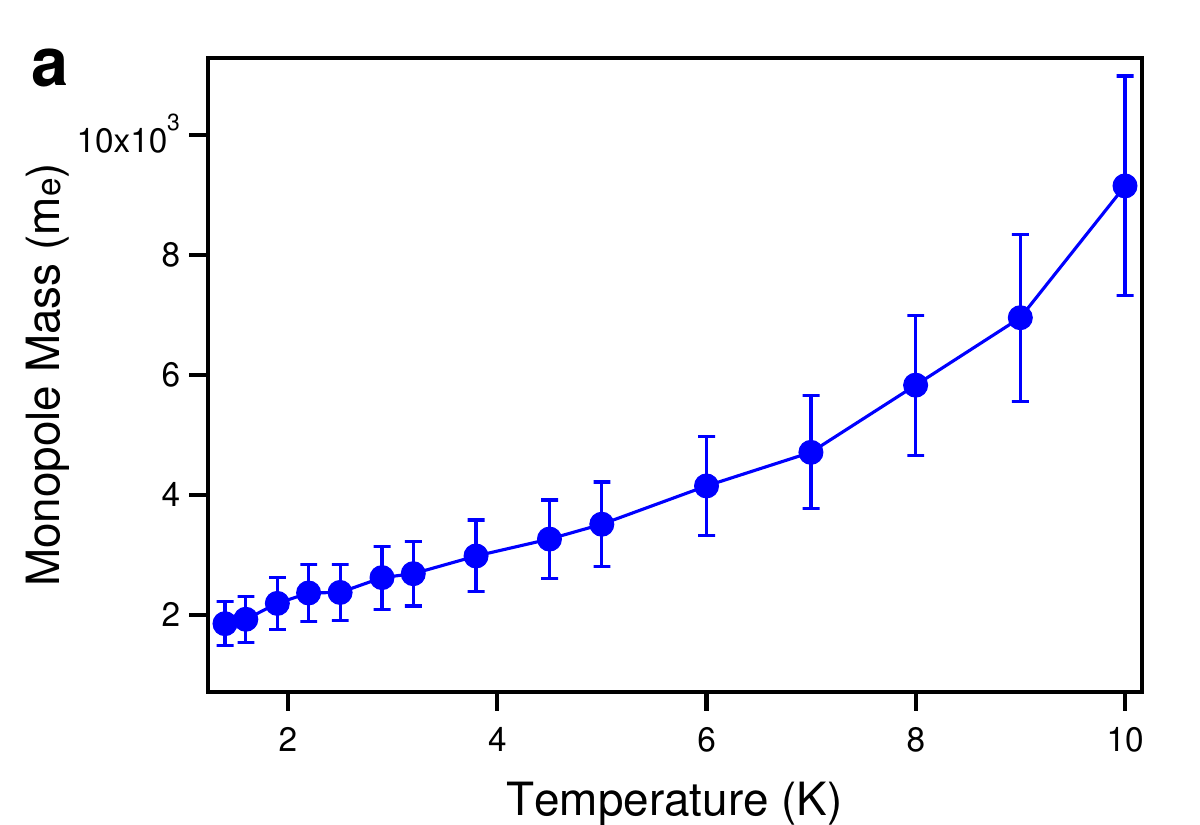}
\includegraphics[trim = 0 5 5 5,width=10cm]{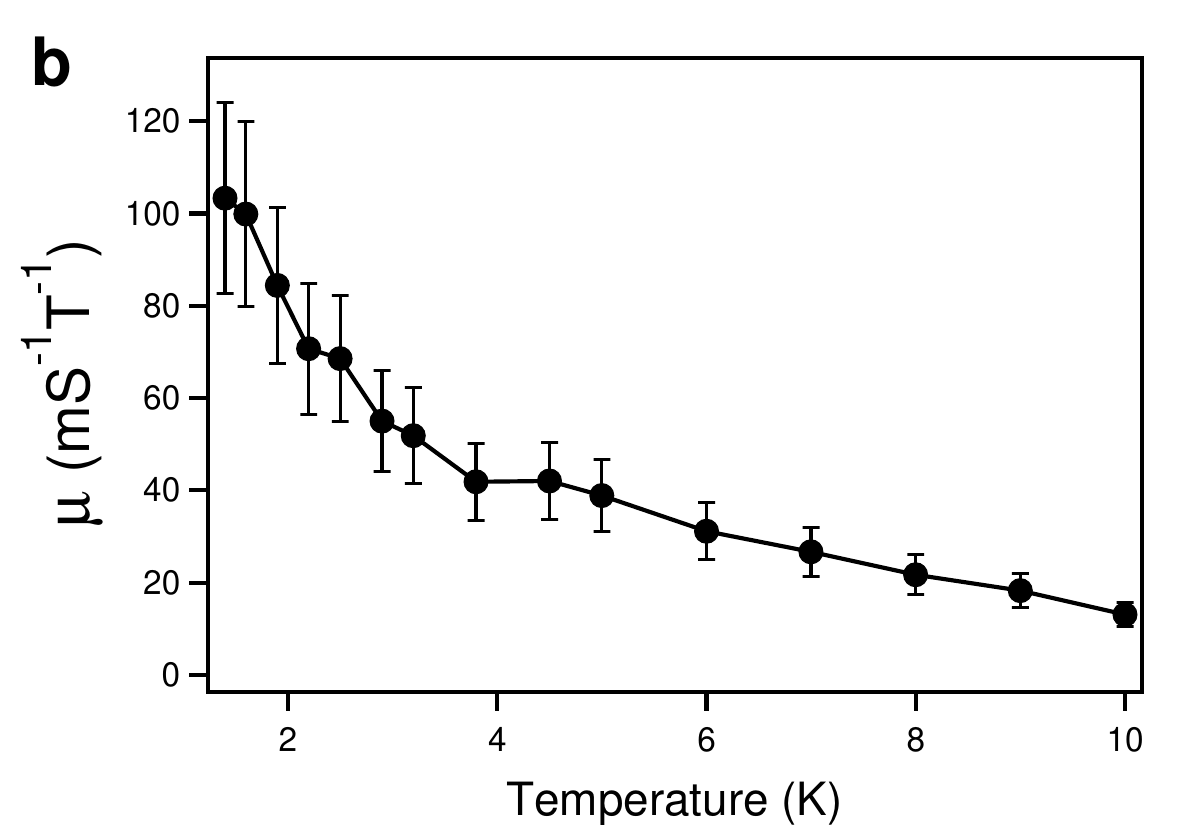}
\label{SI5} 
\caption{ \textsf{\textbf{Temperature dependence of monopole mass and mobility}}  Temperature dependence of (a) the effective monopole mass in units of electron mass, and (b) the monopole mobility.}
\end{figure*}

Following the discussion in the last section, the effective mass and mobility of magnetic monopoles can be obtained from the fitting parameters of the monopole conductivity. The monopole mass can be estimated from the values of the magnetic spectral weight, the monopole charge, as well as the numerical results for the monopole density. The result, in unit of the free electron mass, is plotted in Fig. SI5(a). As discussed in the main text, at low temperatures the values shown here are in good agreement with a low temperature monopole ``band mass" of approximately 2200 $m_e$ calculated in a model with charges hopping on diamond lattice with amplitude $J_{z\pm}$ and using the parameters of Ref. \cite{sRoss11}.  However, at higher temperatures the effective mass increases strongly and becomes of order 10,000 $m_e$ at 10K.   We believe that these large values do not have any physical significance and is an indication that one is no longer in a temperature regime where a monopole model has validity.   At temperature above the mean field T$_c$ the fitting parameters should be considered phenomenological only.

The monopole mobility can be obtained from $\mu = q_m/\gamma m$ once the effective mass is available. The results are shown in Fig. SI5(b). A value several orders of magnitude larger than the ones found in classical spin ice is obtained here \cite{sBovo13}. The huge contrast of monopole mobilities in two different classes of spin ice materials point to essential different physics that governs the low temperature dynamics of the monopole excitations.  As discussed in the main text, although we have performed these fits up to temperature of order 10K, we believe these quantities should only be taken to describe actual physical monopole parameters below the nominal mean-field transition temperature scale of $\sim$ 4K.

\end{widetext}


\begin{thebibliography}{10}

\bibitem{Lacroix11}
Lacroix, C., Mendels, P., Mila, F. (Eds.), Introduction to Frustrated Magnetism, \textit{Springer Series in Solid-State Sciences, Vol. 164}, Springer, 2011.

\bibitem{Balents10}
Balents, L. Spin liquids in frustrated magnets. \textit{Nature}, {\bf{464}}, 199-208 (2010).

\bibitem{Gingras14}
Gingras, M.J.P. \& McClarty, P.A. Quantum spin ice: a search for gapless quantum spin liquids in pyrochlore magnets. \textit{Rep. Prog. Phys.}, {\bf{77}}, 056501 (2014).

\bibitem{Ryzhkin05}
Ryzhkin, I.A. Magnetic relaxation in rare-earth pyrochlores. \textit{J. Exp. Theor. Phys}, {\bf{101}}, 481 -– 486 (2005).

\bibitem{Castelnovo08}
Castelnovo, C., Moessner,R. \& Sondhi, S.L. Magnetic monopoles in spin ice. \textit{Nature} {\bf{451}}, 42 -- 45 (2008).

\bibitem{Ross11a}
Ross, K.A., Savary, L., Gaulin, B.D. \&. Balents, L. Quantum excitations in quantum spin ice. \textit{Phys. Rev. X} {\bf{1}}, 021002 (2011).

\bibitem{Applegate12}
Applegate, R., Hayre, N. R.,Singh, R. R. P., Lin, T., Day, A. G. R., Gingras, M. J. P., Vindication of Yb$_2$Ti$_2$O$_7$ as a Model Exchange Quantum Spin Ice. \textit{Phys. Rev. Lett.} {\bf{109}}, 097205 (2012).

\bibitem{Gardner10}
Gardner, J.S., Gingras, M.J.P. \& Greedan, J.E. Magnetic pyrochlore oxides. \textit{Rev. Mod. Phys.}, {\bf{82}}, 53-107 (2010).

\bibitem{Ehlers03}
Ehlers, G. et al.  Dynamical crossover in ÔhotÕ spin ice.  \textit{J. Phys.: Condens. Matter}  {\bf{15}}   L9--L15 (2003).

\bibitem{Savary13}
Savary, L., Balents, L., Spin liquid regimes at nonzero temperature in quantum spin ice. \textit{Phys. Rev. B} {\bf{87}}, 205130 (2013).

\bibitem{Wan12}
Wan, Y. \& Tchernyshyov, O. Quantum strings in quantum spin ice. \textit{Phys. Rev. Lett.} {\bf{108}}, 247210 (2012).

\bibitem{Hao14}
Hao, Z.H., Day, A.G.R.,\& Gingras, M.J.P., Bosonic many-body theory of quantum spin ice. \textit{Phys. Rev. B} {\bf{90}}, 214430 (2014).

\bibitem{DOrtenzio13}
D'Ortenzio, R.M. et al. Unconventional magnetic ground state in Yb$_2$Ti$_2$O$_7$. \textit{Phys. Rev. B} {\bf{88}}, 134428 (2013).

\bibitem{Pan14}
Pan, L.D., et al. Low-energy electrodynamics of novel spin excitations in the quantum spin ice Yb$_2$Ti$_2$O$_7$. \textit{Nat. Commun.} {\bf{5}}, 4970 (2014).

\bibitem{Hodges01}
Hodges, J.A. et al. The crystal field and exchange interactions in Yb$_2$Ti$_2$O$_7$. \textit{J. Phys.: Condens. Matter} {\bf{13}}, 9301 -- 9310 (2001).

\bibitem{Malkin04}
Malkin, B.Z. et al. Optical spectroscopy of Yb$_2$Ti$_2$O$_7$ and Y$_2$Ti$_2$O$_7$: Yb$^{3+}$ and crystal-field parameters in rare-earth titanate pyrochlores. \textit{Phys. Rev. B} {\bf{70}}, 075112 (2004).

\bibitem{Ross09}
Ross, K.A. et al. Two-dimensional kagome correlations and field induced order in the ferromagnetic xy pyrochlore Yb$_2$Ti$_2$O$_7$. \textit{Phys. Rev. Lett.} {\bf{103}}, 227202 (2009).

\bibitem{Thompson11}
Thompson, J.D. et al. Rods of neutron scattering intensity in Yb$_2$Ti$_2$O$_7$: compelling evidence for significant anisotropic exchange in a magnetic pyrochlore oxide. \textit{Phys. Rev. Lett.} {\bf{106}}, 187202 (2011).

\bibitem{Ross11b}
Ross, K.A. et al. Dimensional evolution of spin correlations in the magnetic pyrochlore Yb$_2$Ti$_2$O$_7$. \textit{Phys. Rev. B} {\bf{84}}, 174442 (2011).

\bibitem{Chang12}
Chang, L.J. et al. Higgs transition from a magnetic Coulomb liquid to a ferromagnet in Yb$_2$Ti$_2$O$_7$. \textit{Nat. Comms.} {\bf{3}}, 992 (2012).

\bibitem{Ross12}
Ross, K.A. et al. Lightly stuffed pyrochlore structure of single-crystalline Yb$_2$Ti$_2$O$_7$ grown by the optical floating zone technique. \textit{Phys. Rev. B} {\bf{86}}, 174424 (2012).

\bibitem{Hayre13}
Hayre, N.R. et al. Thermodynamic properties of Yb$_2$Ti$_2$O$_7$ pyrochlore as a function of temperature and magnetic field: Validation of a quantum spin ice exchange Hamiltonian. \textit{Phys. Rev. B} {\bf{87}}, 184423 (2013).

\bibitem{Vandenborre83}
Vandenborre, M.T. et al. Rare-earth titanates and stannates of pyrochlore structure; vibrational spectra and force fields. \textit{J. Raman Spectrosc.} {\bf{14}}, 63 (1983).

\bibitem{Bovo13}
Bovo, L. et al. Brownian motion and quantum dynamics of magnetic monopoles in spin ice. \textit{Nat. Commun.} {\bf{4}}, 1535 (2013).

\bibitem{Snyder04}
J. Snyder, B. G. Ueland, J. S. Slusky, H. Karunadasa, R. J. Cava, and P. Schiffer. Low-temperature spin freezing in the Dy$_2$Ti$_2$O$_7$ spin ice. \textit{Phys. Rev. B} {\bf{69}}, 064414 (2004).

\bibitem{Jaubert09}
L. D. C. Jaubert and P. C. W. Holdsworth. Signature of magnetic monopole and dirac string dynamics in spin ice. Nature Physics, 5(4):258Ð261, APR 2009.

\bibitem{Gardner04}
J. S. Gardner, G. Ehlers, N. Rosov, R. W. Erwin, and C. Petrovic.  Spin-spin correlations in Yb$_2$Ti$_2$O$_7$: A polarized neutron scattering study.  \textit{Phys. Rev. B} {\bf{70}}, 180404 (2004).

\bibitem{Hohenberg74}
Hohenberg, P.C., \& Brinkman, W.F. Sum rules for the frequency spectrum of linear magnetic chains. \textit{Phys. Rev. B} {\bf{10}}, 128 (1974).

\bibitem{Zaliznyak05}
Zaliznyak, Igor A and Lee, Seung-Hun, Magnetic neutron scattering in Modern Techniques for Characterizing Magnetic Materials, Springer, Heidelberg, 2005.

\bibitem{Onodera93}
Onodera, Y. Breakdown of Debye's model for dielectric relaxation in high frequencies. \textit{J. Phys. Soc. Jpn.} {\bf{62}}, 4104 (1993).


 
\end{thebibliography}

\begin{thebibliography}{10}

\bibitem{sPilon13}
Pilon, D.V., et al. Spin-induced optical conductivity in the spin-liquid candidate Herbertsmithite. \textit{Phys. Rev. Lett.} {\bf{111}}, 127401 (2013).

\bibitem{sVandenborre83}
Vandenborre, M.T. et al. Rare-earth titanates and stannates of pyrochlore structure; vibrational spectra and force fields. \textit{J. Raman Spectrosc.} {\bf{14}}, 63 (1983).

\bibitem{sKlein93} Klein, O., Donovan, S.,  Dressel, M., and Gr\"uner, G., \textit{Int. J. Infrared and Millimeter Waves} \textbf{14} (1993).

\bibitem{sZhai00} Zhai, Z., Kusko, C., Hakim, N., Sridhar, S., Revcolevschi, A., and Vietkine, A., \textit{Review of Scientific Instruments} \textbf{71} 8 (2000).

\bibitem{sPan14}
Pan, L.D., et al. Low-energy electrodynamics of novel spin excitations in the quantum spin ice Yb$_2$Ti$_2$O$_7$. \textit{Nat. Commun.} {\bf{5}}, 4970 (2014).

\bibitem{sRyzhkin05}
Ryzhkin, I.A., Magnetic relaxation in rare-earth pyrochlores. \textit{J. Exp. Theor. Phys}, {\bf{101}}, 481 -? 486 (2005).

\bibitem{sRyzhkin97}
Ryzhkin, I.A. \& Whitworth R.W., The configurational entropy in the Jaccard theory of the electrical properties of ice. \textit{J. Phys: Condens. Matter}, {\bf{9}}, 395 (1997).

\bibitem{sHohenberg74}
Hohenberg, P.C., \& Brinkman, W.F., Sum rules for the frequency spectrum of linear magnetic chains. \textit{Phys. Rev. B} {\bf{10}}, 128 (1974).

\bibitem{sZaliznyak05}
Zaliznyak, Igor A and Lee, Seung-Hun, Magnetic neutron scattering in Modern Techniques for Characterizing Magnetic Materials, Springer, Heidelberg, 2005.

\bibitem{sOnodera93}
Onodera, Y., Breakdown of Debye's model for dielectric relaxation in high frequencies. \textit{J. Phys. Soc. Jpn.} {\bf{62}}, 4104 (1993).


\bibitem{sScalapino93a}
Scalapino, Douglas J., White, Steven R., \& Zhang, Shoucheng, Insulator, metal, or superconductor: The criteria, Phys. Rev. B 47, 7995 (1993).

\bibitem{sAlvarez02a}
AlvarezJ. V., and  Gros, Claudius, Conductivity of quantum spin chains: A quantum Monte Carlo approach, Phys. Rev. B 66, 094403 (2002).

\bibitem{sCoffey04}
Coffey, William T.,  Dielectric relaxation: an overview. Journal of Molecular Liquids {\bf{114}}, 5-25 (2004).

\bibitem{sRoccard33a}
Rocard, Y., Analyse des orientations mol\'eculaires de mol\'ecules \`a moment permanent dans un champ alternatif. Application \`a la dispersion de la constante di\'electrique et \`a l'effet Kerr.  J. Phys. Radium 4, 247-250 (1933).

\bibitem{sGross55} Gross, E. P., Inertial effects and dielectric relaxation,  J. Chem. Phys. 23, 1415 (1955).

\bibitem{sSack57}  Sack, R. A., Relaxation Processes and Inertial Effects I: Free Rotation about a Fixed Axis,
Proc. Phys. Soc. London, Sect. B 70, 402 (1957).

\bibitem{sRoss11}
Ross, K.A., Savary, L., Gaulin, B.D. \&. Balents, L. Quantum excitations in quantum spin ice. \textit{Phys. Rev. X} {\bf{1}}, 021002 (2011).

\bibitem{sBovo13}
Bovo, L. et al. Brownian motion and quantum dynamics of magnetic monopoles in spin ice. \textit{Nat. Commun.} {\bf{4}}, 1535 (2013).


\end{thebibliography}
\end{document}